\documentclass[prb,twocolumn,showpacs,preprintnumbers,amsmath,amssymb]{revtex4}
\usepackage{graphicx}% Include figure files
\usepackage{dcolumn}% Align table columns on decimal point
\usepackage{bm}% bold math
\usepackage{latexsym,epsfig}

\begin{document}
\preprint{08 January 2012}

\title{Bose-Hubbard Models in Confining Potentials: An Inhomogeneous
Mean-Field Theory}

\author{Ramesh V. Pai}
\email{rvpai@unigoa.ac.in} \affiliation{Department of Physics, Goa
University, Taleigao Plateau, Goa 403 206, India}
\author{Jamshid Moradi Kurdestany}
\email{jamshid@physics.iisc.ernet.in} \affiliation{Centre for
Condensed Matter Theory, Department of Physics, Indian Institute of
Science, Bangalore 560 012, India}
\author{K. Sheshadri}
\email{kshesh@gmail.com} \affiliation{Bagalur, Bangalore North
Taluk, India 562 149, India}
\author{Rahul Pandit}
\email{rahul@physics.iisc.ernet.in} \altaffiliation[Also
at~]{Jawaharlal Nehru Centre For Advanced Scientific Research,
Jakkur, Bangalore, India} \affiliation{Centre for Condensed Matter
Theory, Department of Physics, Indian Institute of Science,
Bangalore 560012, India.}

\date{\today}
\begin{abstract}
We present an extensive study of Mott insulator (MI) and superfluid
(SF) shells in Bose-Hubbard (BH) models for bosons in optical
lattices with harmonic traps. For this we develop an inhomogeneous
mean-field theory. Our results for the BH model with one type of
spinless bosons agrees quantitatively with quantum Monte Carlo (QMC)
simulations. Our approach is numerically less intensive than such
simulations, so we are able to perform calculation on experimentally
realistic, large 3D systems, explore a wide range of parameter
values, and make direct contact with a variety of experimental
measurements. We also generalize our inhomogeneous mean-field theory
to study BH models with harmonic traps and (a) two species of bosons
or (b) spin-1 bosons. With two species of bosons we obtain rich
phase diagrams with a variety of SF and MI phases and associated
shells, when we include a quadratic confining potential. For the
spin-1 BH model we show, in a representative case, that the system
can display alternating shells of polar SF and MI phases; and we
make interesting predictions for experiments in such systems.
\end{abstract}

\maketitle

\section{Introduction}

High-precision experiments on cold atoms, such as spin-polarized
$^{87}$Rb, in traps have provided powerful methods for the study of
quantum phase transitions~\cite{rmp}, e.g., the transition from a
superfluid (SF) to a bosonic Mott-insulator (MI) in an optical
lattice~\cite{jaksch,greiner}. This transition was predicted by
mean-field studies~\cite{fisher,sheshadri} and obtained in
Monte-Carlo simulations~\cite{mc} of the Bose-Hubbard model before
it was seen in experiments~\cite{rmp,jaksch,greiner}.  Recent
experiments~\cite{catani08,trotzky08} have investigated a
heteronuclear degenerate mixture of two bosonic species, e.g.,
$^{87}$Rb and $^{41}$K, in a three-dimensional optical lattice; such
mixtures have also been studied
theoretically~\cite{kuklov03,han04,buonsante08,hu09,ozaki09} and by
Monte Carlo simulations~\cite{roscilde07}. Systems of alkali atoms
with nuclear spin $I=3/2$ have hyperfine spin $F=1$; examples
include $^{23}$Na, $^{39}$K, and $^{87}$Rb; these spins are frozen
in magnetic traps, so these atoms are treated as spinless bosons;
however, in purely optical traps, such spins can form spinor
condensates~\cite{spin1expt,tlho,mukerjee,rvpspin1}. Thus, we
consider the following three types of Bose-Hubbard (BH) models: (1)
a BH model for spinless interacting bosons of one type; (2) a
generalization of the spinless BH model with two types of bosons;
and (3) a spin-1 generalization of the spinless BH model with bosons
of one type. We study these models by developing extensions of an
inhomogeneous mean-field theory~\cite{sheshprl}, which has been used
for the Bose-glass phase in the disordered BH model.

In addition to the optical-lattice potential, a confining potential,
which is typically quadratic, is present in all experiments. This
inhomogeneous potential leads to inhomogeneities in the phases that
are obtained: simulations~\cite{wessel,svistunov} of the
Bose-Hubbard model with this confining potential and
experiments~\cite{bloch,campbell} on interacting bosons in optical
lattices with a confining potential have both seen alternating
shells of SF and MI regions in the single-species, spinless case. We
explore such shells via the inhomogeneous mean-field theory, first
for single-species, spinless bosons and then for the two-species and
spin-1 generalizations mentioned above.

Mean-field theories for the Bose-Hubbard model were first developed
for the homogeneous case~\cite{fisher,sheshadri}; these theories
were then extended to the inhomogeneous case~\cite{sheshprl} to
develop an understanding of the Bose-glass phase in the disordered
Bose-Hubbard model. We show that BH models with  confining potential
can be treated, at the level of mean-field theory, as was done in
the Bose-glass case~\cite{sheshprl}; in particular, we provide a
natural framework for understanding alternating SF and MI shells,
which are seen in simulations~\cite{wessel,svistunov} and
experiments~\cite{bloch,campbell} on interacting bosons, trapped in
a confining potential, and in an optical lattice. Though other
groups~\cite{bergkvist,pollet,demarco05,mitra08,spiel,kato08} have
studied such shell structure theoretically, they have not obtained
the quantitative agreement with quantum Monte Carlo (QMC)
simulations~\cite{wessel} that we obtain, except in one
dimension~\cite{batrouni08}. Furthermore, our theory yields results
in good agreement with a variety of experiments; and it can be
generalized easily to (a) two species of interacting bosons and (b)
the spin-$S$ case, as we show explicitly for $S=1$; in both these
cases we provide interesting predictions that will, we hope,
stimulate new experiments. Our inhomogeneous mean-field calculations
can be carried out with experimentally realistic parameters, so we
can make direct comparison with experiments. In particular, we
obtain in-trap density distributions of alternating SF and MI
shells; these show plateaux in certain regions, which can be
understood on the basis of simple geometrical arguments.
Furthermore, we obtain the radii of SF and MI shells from in-trap
density distributions and demonstrate how the phase diagram of the
homogeneous Bose-Hubbard model can be obtained from these radii. We
also obtain results that are of direct relevance to recent
atomic-clock-shift experiments~\cite{campbell}. With two species of
bosons we obtain phase diagrams in the homogeneous case over a far
wider range of parameters than has been reported hitherto. We find
rich phase diagram with phases that include ones in which (a) both
types of bosons are in SF states, (b) both types of bosons are in MI
phases with different or the same densities, and (c) one type of
boson is in an SF phase whereas the other type is in an MI phase. We
show that each of these phases appear in shells when we include a
quadratic confining potential; and we also obtain in-trap density
distributions that shows plateaux as in the single-species case. In
the case of the spin-1 Bose-Hubbard model we show, in a
representative case, that the system can display alternating shells
of polar SF~\cite{rvpspin1} and MI phases; the latter have integral
values for the boson density. Our inhomogeneous mean-field theory
leads to interesting predictions for atomic-clock-shift experiments
in systems with spin-1 bosons in an optical lattice with a confining
potential.

The remaining part of this paper is organized as follows. In Sec.
2 we describe the models we use and how we develop an
inhomogeneous mean-field theory for them. Section 3 is devoted to
our results; subsection 3A contains the results of our
inhomogeneous mean-field theory for the single-species, spinless
Bose-Hubbard model; subsection 3B is devoted to the results, for
both homogeneous and inhomogeneous cases, for the spinless BH
model with two types of bosons; subsection 3C is devoted to our
results for the single-species BH model for spin-1 bosons.
Section 4 contains our conclusions, a comparison of our work with
earlier studies in this area, and the experimental implications
of our results.

\section{Models and inhomogeneous mean-field theory}

We begin by defining the three Bose-Hubbard models that we study. We then
develop inhomogeneous mean-field theories that are well suited for studying
the spatial organization of phases in these models with confining potentials.

\subsection{Models}

The simplest Bose-Hubbard model describes a single species of
spinless bosons in an optical lattice by the following
Hamiltonian:
\begin{equation}
\frac{{\cal H}}{zt} = -\frac{1}{z} \sum_{<i,j>}
(a_{i}^{\dagger} a_{j} + h.c)+\frac{1}{2}\frac{U}{zt} \sum_{i} {\hat
n}_{i} ({\hat n}_{i} -1) - \frac{1}{zt}\sum_{i} \mu_i{\hat n}_{i};
\label{eq:bh0}
\end{equation}
here spinless bosons hop between the $z$ nearest-neighbor pairs of
sites $<i,j>$ with amplitude $t$, $a_{i}^\dagger,\, a_{i},$ and
$\hat{n_{i}}\equiv a^{\dagger}_{i}a_{i}$ are, respectively, boson
creation, annihilation, and number operators at the sites $i$ of a
$d$-dimensional hypercubic lattice (we study $d=2$ and $3$), $U$ the
onsite Hubbard repulsion, $\mu_i \equiv \mu-V_T R^2_i$, $\mu$ the
uniform chemical potential that controls the total number of bosons,
$V_T$ the strength of the harmonic confining potential, and
$R^2_i\equiv \sum_{n=1}^d X^2_n(i)$, where $X_n(i),\, 1\leq n \leq
d$, are the Cartesian coordinates of the site $i$ (in $d=3$ $X_1 =
X$, $X_2 = Y$, and $X_3 = Z$) ; the origin is chosen to be at the
center of the lattice.  In terms of experimental
parameters~\cite{rmp} $\frac{U}{zt}=\frac{\sqrt{8}\pi}{4z}
\frac{a_s}{a}e^{2\sqrt{\frac{V_0}{E_r}}}$, where $E_r$ is the recoil
energy, $V_0$ the strength of the lattice potential, $a_s$
($=5.45$nm for $^{87}$Rb) the s-wave scattering coefficient,
$a=\lambda/2$ the optical lattice constant, and $\lambda=825$nm the
wavelength of the laser used to create the optical lattice;
typically $0 \leq V_0 \leq 22 E_r$. We set the scale of energies by
using $zt = 1$ in the Bose-Hubbard model~\ref{eq:bh0}; for
comparisons with experimental systems we should scale all energies
by $E_r$.

For a mixture with two types of bosons, we use the following Bose-Hubbard
Hamiltonian:
\begin{eqnarray}
\frac{{\cal H}}{z} &=& -\frac{t_a}{z} \sum_{<i,j>} (a_{i}^{\dagger}
a_{j} + h.c)-\frac{t_b}{z} \sum_{<i,j>} (b_{i}^{\dagger} b_{j} +
h.c)\\ \nonumber &&+\frac{1}{2}\frac{U_a}{z} \sum_{i} {\hat n}_{ai}
({\hat n}_{ai} -1)+\frac{1}{2}\frac{U_b}{z} \sum_{i} {\hat n}_{bi}
({\hat n}_{bi} -1)\\ \nonumber &&+\frac{U_{ab}}{z} \sum_{i} {\hat
n}_{ai}{\hat n}_{bi}-\frac{1}{z}\sum_{i} \mu_{ai}{\hat n}_{ai}
-\frac{1}{z}\sum_{i} \mu_{bi}{\hat n}_{bi}; \label{eq:bhab}
\end{eqnarray}
the first and second term represent, respectively, the hopping of
bosons of types $a$ and $b$ between the nearest-neighbor pairs of
sites $<i,j>$ with hopping amplitudes $t_a$ and $t_b$; here
$a_{i}^\dagger,\, a_{i},$ and ${\hat n}_{ai}\equiv
a^{\dagger}_{i}a_{i}$ and $b_{i}^\dagger,\, b_{i},$ and ${\hat
n}_{bi}\equiv b^{\dagger}_{i}b_{i}$ are, respectively, boson
creation, annihilation, and number operators at the sites $i$ of a
$d$-dimensional hypercubic lattice (we study $d=2$ and $3$) for the
two bosonic species.  For simplicity we restrict ourselves to the
case $t_a = t_b = t$, and, to set the scale of energy, we use $zt =
1 $. The third and fourth terms account for the onsite interactions
of bosons of a given type, with energies $U_a$ and $U_b$,
respectively, whereas the fifth term, with energy $U_{ab}$, arises
because of the onsite interactions between bosons of types $a$ and
$b$.  We have two chemical potential terms: $\mu_{ai} \equiv
\mu_{a}-V_{Ta} R^2_i$ and $\mu_{bi} \equiv \mu_{b}-V_{Tb} R^2_i$,
where $\mu_a$ and $\mu_b$ the uniform chemical potentials that
control the total number of bosons, of species $a$ and $b$,
respectively, $V_{Ta}$ and  $V_{Tb}$ are the strengths of their
harmonic confining potentials (we restrict ourselves to the case
$V_{Ta} = V_{Tb}$), and  $R^2_i\equiv \sum_{n=1}^d X^2_n(i)$, where
$X_n(i),\, 1\leq n \leq d$, are the Cartesian coordinates of the
site $i$; the origin is at the center of the lattice.

The spin-1 Bose-Hubbard Hamiltonian~\cite{rvpspin1} that we consider is
\begin{eqnarray}
{\frac{\cal{H}}{zt}}&=&-{\frac{1}{z}}
\sum_{<i,j>,\sigma}(a^{\dagger}_{i,\sigma}a_{j,\sigma}+ h.c) +
\frac{1}{2}\frac{U_0}{zt} \sum_i {\hat{n}}_i({\hat{n}}_i-1)
\nonumber \\
&&+\frac{1}{2}\frac{U_2}{zt} \sum_i (\vec{F}^2_i -
2{\hat{n}}_i)-\frac{1}{zt}\sum_{i} \mu_i{\hat{n}}_i.
\label{eq:bhspin1}
\end{eqnarray}
Here spin-1 bosons can occupy the sites $i$ of a $d-$dimensional, hypercubic
lattice and hop between the $z$ nearest-neighbor pairs of sites $<i,j>$ with
amplitude $t$, $\sigma$ is the spin index that can be $1,0,-1$,
$a^{\dagger}_{i,\sigma}$ and $a_{i,\sigma}$ are, respectively, site- and
spin-dependent boson creation and annihilation operators, and the number
operator ${\hat{n}}_{i\sigma}\equiv a^{\dagger}_{i,\sigma}a_{i,\sigma}$; the
total number operator at site $i$ is
${\hat{n}}_i\equiv\sum_\sigma{\hat{n}}_{i,\sigma}$, and
$\vec{F}_i=\sum_{\sigma,\sigma '} a^{\dagger}_{i,\sigma} \vec{F}_{\sigma
,\sigma '} a_{i,\sigma '}$ with $ \vec{F}_{\sigma ,\sigma '}$ standard spin-1
matrices. The model~(\ref{eq:bhspin1}) includes, in addition to the onsite
repulsion $U_0$, an energy $U_2$ for nonzero spin configurations on a site.
Such a spin-dependent term arises from the difference between the
scattering lengths for $S=0$ and $S=2$ channels~\cite{law}. The inhomogeneous
chemical potential $\mu_i$ is related to the uniform chemical potential $\mu$
and the quadratic, confining potential as in the spinless case~\ref{eq:bh0}.
We set the scale of energies by choosing $zt = 1$.

\subsection{Inhomogeneous mean-field theory}

The mean-field theory we use has been very successful in obtaining the phase
diagrams for models~(\ref{eq:bh0}) and (\ref{eq:bhspin1}), with $V_T = 0$,
i.e., in the absence of the harmonic confining
potential~\cite{sheshadri,rvpspin1}. The inhomogeneous generalization of this
theory, developed first for the Bose-glass phase~\cite{sheshprl} and spinless
bosons, decouples the hopping term to obtain an effective one-site problem,
neglects quadratic deviations from equilibrium values (denoted by angular
brackets), uses the approximation
\begin{equation}
\label{eq:hop} a^{\dagger}_{i}a_{j} \simeq \langle
a^{\dagger}_{i}\rangle a_{j} +a^{\dagger}_{i}\langle a_{j}\rangle
-\langle a^{\dagger}_{i}\rangle \langle a_{j}\rangle ,
%\label{eq:hop0}
\end{equation}
introduces the superfluid order parameter $\psi_{i}\equiv \langle
a_{i}\rangle$ for the site $i$, and thence expresses the
Hamiltonian~(\ref{eq:bh0}) as ${\cal{H}}^{MF}=\sum_{i} {\cal{H}}_i^{MF}$,
where the superscript $MF$ denotes mean field and the single-site Hamiltonian
is
\begin{equation}
 \frac{{\cal{H}}_i^{MF}}{zt}=\frac{1}{2}\frac{U}{zt}
 {\hat{n}}_i({\hat{n}}_i-1)
-\frac{\mu_i}{zt} {\hat{n}}_i -(\phi_ia^{\dagger}_{i}+\phi_i^*a_{i})+
\psi_i^* \phi_i.
\label{eq:mfh0}
\end{equation}
Here $\phi_i\equiv \frac{1}{z}\sum_{\delta} \psi_{i+\delta}$ and $\delta$
labels the $z$ nearest neighbors of the site $i$. If $V_T=0$, the effective
onsite chemical potential $\mu_i=\mu$, for all $i$, so the local density and
superfluid order parameters are independent of $i$: $\rho_i=\rho$ and
$\psi_i=\psi$. If $V_T > 0$ we first obtain the matrix elements of
${\cal{H}}_i^{MF}$ in the onsite, occupation-number basis $\{|n_i\rangle\}$,
truncated in practice by choosing a finite value for $n_{\mbox{max}}$, the
total number of bosons per site, for a given initial set of values for
$\{\psi_i\}$.  [For small values of $U$ we must use large values of
$n_{\mbox{max}}$; for the values of $U$ we consider $n_{\mbox{max}}=6$
suffices.] We then diagonalize this matrix, which depends on $\psi_i$ and
$\psi_{i+\delta}$, to obtain the lowest energy and the corresponding wave
function, denoted, respectively, by $E_g^i(\psi_i,\psi_{i+\delta})$ and
$\Psi_g(\{\psi_i\})$; from these we obtain the new superfluid order parameters
$\psi_i = \langle \Psi_g(\{\psi_i\})\mid a_i \mid \Psi_g(\{\psi_i\})
\rangle$; we use these new values of $\psi_i$ as inputs to reconstruct
${\cal{H}}_i^{MF}$ and repeat the diagonalization procedure until we achieve
self consistency of input and output values to obtain the equilibrium value
$\psi^{eq}_i$ (henceforth we suppress the superscript $eq$ for notational
convenience). [This is equivalent to a minimization of the total energy
$E_g(\{\psi_i\})\equiv\sum_i E_g^i(\psi_i,\psi_{i+\delta})$ with respect to
$\psi_i$; if more than one solution is obtained, we pick the one that yields
the global minimum.] The onsite density is obtained from $\rho_i=\langle
\Psi_g(\{\psi_i\})\mid {\hat{n}}_i \mid \Psi_g(\{\psi_i\})
\rangle$. In representative cases, we have found that the equilibrium
value of $\psi_i$ is real; so, henceforth, we restrict ourselves
to real values of $\psi$.

For the two-species Hamiltonian~(\ref{eq:bhab}) our mean-field
theory obtains an effective one-site problem by decoupling the two
hopping terms as follows (cf., Eq.~\ref{eq:hop}):
\begin{eqnarray}
a^{\dagger}_{i}a_{j} &\simeq& \langle a^{\dagger}_{i}\rangle
a_{j} +a^{\dagger}_{i}\langle a_{j}\rangle -\langle a^{\dagger}_{i}\rangle
\langle a_{j}\rangle ; \nonumber \\
b^{\dagger}_{i}b_{j} &\simeq& \langle b^{\dagger}_{i}\rangle
b_{j} +b^{\dagger}_{i}\langle b_{j}\rangle -\langle b^{\dagger}_{i}\rangle
\langle b_{j}\rangle ;
\label{eq:hopab}
\end{eqnarray}
here the superfluid order parameters for the site $i$ for
bosons of types $a$ and $b$ are
$\psi_{{ai}}\equiv \langle a_{i}\rangle$ and $\psi_{{bi}}\equiv \langle b_{i}\rangle$, respectively.
The approximation~(\ref{eq:hopab}) can now be used to write the
Hamiltonian~(\ref{eq:bhab}) as a sum over single-site, mean-field
Hamiltonians ${\cal{H}}_i^{MF}$ (cf., Eq.~\ref{eq:mfh0}) given below:
\begin{eqnarray}
 \frac{{\cal{H}}_i^{MF}}{zt}&=&\frac{1}{2}\frac{U_a}{zt}
 {\hat{n}}_{ai}({\hat{n}}_{ai}-1)
-\frac{\mu_{ai}}{zt} {\hat{n}}_{ai}\\ \nonumber
&&-(\phi_{ai}a^{\dagger}_{i} +\phi_{ai}^*a_{i})+ \psi_{ai}^*
\phi_{ai}\\ \nonumber &&+\frac{1}{2}\frac{U_b}{zt}
{\hat{n}}_{bi}({\hat{n}}_{bi}-1) -\frac{\mu_{bi}}{zt}
{\hat{n}}_{bi}\\ \nonumber
&&-(\phi_{bi}b^{\dagger}_{i}+\phi_{bi}^*b_{i})+ \psi_{bi}^*
\phi_{bi}+\frac{U_{ab}}{zt} {\hat{n}}_{ai}{\hat{n}}_{bi}.
\label{eq:mfhab}
\end{eqnarray}

Here $\phi_{ai}\equiv \frac{1}{z}\sum_{\delta} \psi_{{ai}+\delta}$ and
$\phi_{bi}\equiv \frac{1}{z}\sum_{\delta} \psi_{{bi}+\delta}$, where $\delta$
labels the nearest neighbors of the site $i$.  If $V_T=0$, the effective
onsite chemical potentials $\mu_{ai}=\mu_{a}$ and $\mu_{bi}=\mu_{b}$, for all
$i$, so $\rho_{ai}=\rho_{a}$, $\rho_{bi}=\rho_{b}$, $\psi_{ai}=\psi_{a}$, and
$\psi_{bi}=\psi_{b}$ are independent of $i$.

If $V_T > 0$, we first obtain, for a given initial set of values for
$\{\psi_{ai}\}$ and $\{\psi_{bi}\}$, the matrix elements of
${\cal{H}}_i^{MF}$ in the onsite, occupation-number basis
$\{|n_{ai}\rangle\ , |n_{bi}\rangle\}$, which we truncate in a
practical calculation by choosing a finite value $n_{\mbox{max}}$
for the total number of bosons per site. [The smaller the values of
the interaction parameters $U_a, U_b$, and $U_{ab}$ the larger must
be the value of $n_{\mbox{max}}$; for the values of $U_a ,\ U_b , \
U_{ab},\ \mu_a$, and $\mu_b$ that we consider, $n_{\mbox{max}}=6$
suffices.] We then diagonalize this matrix, which depends on
$\psi_{ai}$ , $\psi_{bi}$, $\psi_{a(i+\delta)}$, and
$\psi_{b(i+\delta)}$ to obtain the lowest energy and the
corresponding wave function, denoted, respectively, by
$E_g^i(\psi_{ai},\psi_{a(i+\delta)};\psi_{bi},\psi_{b(i+\delta)})$
and $\Psi_g(\{\psi_{ai},\psi_{bi}\})$, whence we obtain the new
superfluid order parameters $\psi_{ai} = \langle
\Psi_g(\{\psi_{ai},\psi_{bi}\})\mid a_i \mid
\Psi_g(\{\psi_{ai},\psi_{bi}\}) \rangle$ and $\psi_{bi} = \langle
\Psi_g(\{\psi_{ai},\psi_{bi}\})\mid b_i \mid
\Psi_g(\{\psi_{ai},\psi_{bi}\}) \rangle$; we use these new values of
$\psi_{ai}$ and $\psi_{bi}$ as inputs to reconstruct
${\cal{H}}_i^{MF}$ and repeat the diagonalization procedure until we
achieve self consistency of input and output values to obtain the
equilibrium value $\psi^{eq}_{ai}$ and $\psi^{eq}_{bi}$; again we
suppress the superscript $eq$ for notational convenience. [As we
have mentioned in the single-species case, this self-consistency
procedure is equivalent to a minimization,  with respect to
$\psi_{ai}$ and $\psi_{bi}$, of the total energy
$E_g(\{\psi_{ai},\psi_{bi}\})\equiv\sum_i
E_g^i(\psi_{ai},\psi_{{ai}+\delta};\psi_{bi},\psi_{{bi}+\delta})$;
we pick the one that yields the global minimum.] The onsite
densities are obtained from $\rho_{ai}=\langle
\Psi_g(\{\psi_{ai},\psi_{bi}\})\mid {\hat{n}}_{ai} \mid
\Psi_g(\{\psi_{ai},\psi_{bi}\}) \rangle$ and $\rho_{bi}=\langle
\Psi_g(\{\psi_{ai},\psi_{bi}\})\mid {\hat{n}}_{bi} \mid
\Psi_g(\{\psi_{ai},\psi_{bi}\}) \rangle$, respectively. We follow
our discussion of the mean-field theory of the BH
model~(\ref{eq:bh0}) and restrict ourselves to real values of
$\psi_{ai}$ and $\psi_{bi}$.

The inhomogeneous mean-field theory for the spin-1 BH model follows
along similar lines. The spin-1 analogs of
Eqs.~\ref{eq:hop} and \ref{eq:mfh0} are respectively,
\begin{eqnarray}
a^{\dagger}_{i,\sigma}a_{j,\sigma} &\simeq&
\langle a^{\dagger}_{i,\sigma}\rangle a_{j,\sigma} +
a^{\dagger}_{i,\sigma}\langle a_{j,\sigma}\rangle -
\langle a^{\dagger}_{i,\sigma}\rangle \langle a_{j,\sigma}\rangle
\label{eq:hopspin1}
\end{eqnarray}
and
\begin{eqnarray}
\nonumber  \frac{{\cal{H}}_i^{MF}}{zt}&=&\frac{1}{2}\frac{U_0}{zt}
{\hat{n}_i}({\hat{n}_i}-1) + \frac{1}{2}\frac{U_2}{zt} ({\vec
F}^2_i-2{\hat{n}_i}) -\frac{\mu_i}{zt} {\hat{n}_i}\\
&&-\sum_\sigma(\phi_{i,\sigma}a^{\dagger}_{i,\sigma}+
\phi_{i,\sigma}^*a_{i,\sigma})+ \sum_\sigma\psi_{i,\sigma}^*
\phi_{i,\sigma}. \label{eq:mfhspin1}
\end{eqnarray}
Here we use the following superfluid order parameters:
\begin{equation}
\psi_{i,\sigma}\equiv \langle a_{i,\sigma}\rangle;
\label{eq:opspin1}
\end{equation}
and $\phi_{i,\sigma}\equiv \frac{1}{z}\sum_{\delta}
\psi_{(i+\delta),\sigma}$, where and $\delta$ labels the $z$ nearest
neighbors of the site $i$; recall, furthermore, that $\sigma$ can
assume the values $1,\,0,\,-1$, and ${\hat{n}}_{i,\sigma}\equiv
a^{\dagger}_{i,\sigma}a_{i,\sigma}$,
${\hat{n}}_i\equiv\sum_\sigma{\hat{n}}_{i,\sigma}$, and
$\vec{F}_i=\sum_{\sigma,\sigma '} a^{\dagger}_{i,\sigma}
\vec{F}_{\sigma ,\sigma '} a_{i,\sigma '}$ with $ \vec{F}_{\sigma
,\sigma '}$ standard spin-1 matrices. With these order parameters~
(cf., Eq.~\ref{eq:opspin1}) we have developed an inhomogeneous
version of the homogeneous mean-field theory~\cite{rvpspin1} for the
spin-1 BH model with $V_T = 0$.

The self-consistency procedure that we use now is similar to, but
more complicated than, the one we have used for the spinless BH
model.  If $V_T > 0$ we first obtain, for a given initial set of
values for $\{\psi_{i,\sigma}\}$, the matrix elements of
${\cal{H}}_i^{MF}$ in the onsite, occupation-number basis
$\{|n_{i,-1}, n_{i,0},n_{i,1}\rangle\}$, truncated in a practical
calculation by choosing a finite value for $n_{\mbox{max}}$, the
total number of bosons per site,  [For small values of $U$ and $U_2$
we must use large values of $n_{\mbox{max}}$; for the values we use
here, $n_{\mbox{max}}=4$ suffices.] We then diagonalize this matrix,
which depends on $\psi_{i,\sigma}$ and $\psi_{(i+\delta),\sigma}$,
to obtain the lowest energy and the corresponding wave function,
denoted, respectively, by
$E_g^i(\psi_{i,\sigma},\psi_{(i+\delta),\sigma})$ and
$\Psi_g(\{\psi_{i,\sigma}\})$; from these we obtain the new
superfluid order parameters $\psi_{i,\sigma} = \langle
\Psi_g(\{\psi_{i,\sigma}\})\mid a_{i,\sigma} \mid
\Psi_g(\{\psi_{i,\sigma}\}) \rangle$; we use these new values of
$\psi_{i,\sigma}$ to reconstruct ${\cal{H}}_i^{MF}$ and repeat the
diagonalization procedure until input and output values are self
consistent; thus we obtain the equilibrium value
$\psi^{eq}_{i,\sigma}$. We suppress $eq$ as above and recall that
this self-consistent procedure is equivalent to a minimization of
the total energy in the spin-1 case~\cite{rvpspin1}. Here too, we
follow our discussion of the mean-field theory of the BH
model~(\ref{eq:bh0}) and restrict ourselves to real values of
$\psi_{i,\sigma}$.

We have noted in an earlier study~\cite{rvpspin1} that, at the level of our
mean-field theory, the superfluid density in the spin-1 case is
\begin{equation}
\label{eq:rho_s}
\rho_s=\sum_\sigma \mid \psi_\sigma^{eq} \mid^2;
\end{equation}
and the magnetic properties of the SF phases follow
from~\cite{tlho,mukerjee}
\begin{equation}
\langle\vec{F}\rangle=\frac{\sum_{\sigma,\sigma '}
\psi_{\sigma}^{eq} \vec{F}_{\sigma ,\sigma '} \psi_{\sigma '}^{eq}}
{\sum_{\sigma}|\psi_\sigma^{eq}|^2}.
\end{equation}
If we substitute the explicit forms of the spin-1 matrices we
obtain
\begin{eqnarray}
\label{eq:s_ave}
\langle\vec{F}\rangle &=& \sqrt{2}\frac{( \psi_1 \psi_0 + \psi_{-1}
\psi_0)} {\sum_{\sigma}|\psi_\sigma|^2} \hat{x} +
\frac{(\psi_1^2-\psi_{-1}^2)}{\sum_{\sigma}|\psi_\sigma|^2} \hat{z},
\nonumber \\
\langle\vec{F}\rangle^2 &=& 2 \frac{( \psi_1 \psi_0 + \psi_{-1}
\psi_0)^2} {(\sum_{\sigma}|\psi_\sigma|^2)^2}
+\frac{(\psi_1^2-\psi_{-1}^2)^2}{(\sum_{\sigma}|\psi_\sigma|^2)^2},
\end{eqnarray}
where $\hat{x}$ and $\hat{z}$ are unit vectors in spin space;
SF phases with $\langle\vec{F}\rangle=0$ and
$\langle\vec{F}\rangle^2=1$ are referred to as polar and
ferromagnetic, respectively. The order-parameter manifolds of
these phases can be found in earlier studies~\cite{mukerjee,rvpspin1}.

\section{Results}

Given the formalism we have developed above, we can obtain several results
for quantities that have been measured in quantum Monte Carlo (QMC)
simulations or in experiments for the spinless case.  We cover this in
Subsection 3A. Subsection 3B is devoted to the results of our inhomogeneous
MF theory for the case with two types of bosons. Subsection 3C is devoted to
the results of our inhomogeneous MF theory for the spin-1 case.

\begin{figure}[htbp]
\centering \epsfig{file=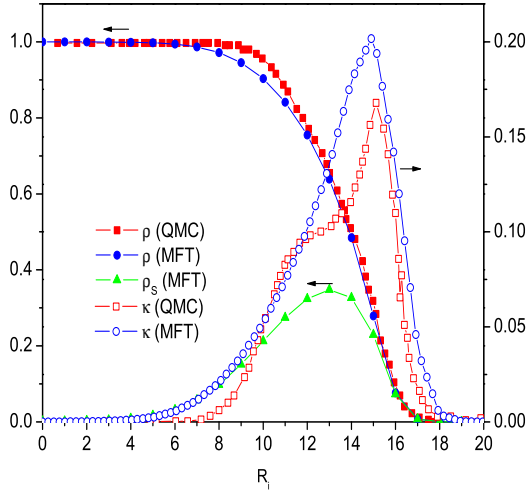,width=8cm,height=8cm}
\caption{(Color online) Plots with a comparison of our MF results
for the local density $n_i$ (blue filled circles), local
compressibility $\kappa^{local}_i= \partial n/\partial \mu^{eff}$
(blue open circles), and local superfluid density
$\rho^s_i=\psi^2_i$ (green filled triangles) of spinless bosons in a
two-dimensional parabolic trap with $V_T/U=0.002$, $\mu/U=0.37$ and
$U/t=25$; we have obtained QMC data by digitizing plots in figures
in simulation studies~\cite{wessel} for $n_i$ (red filled squares)
and $\kappa^{local}_i$ (red open squares).} \label{fig:qmc}
\end{figure}

\begin{figure}[htbp]
\centering \epsfig{file=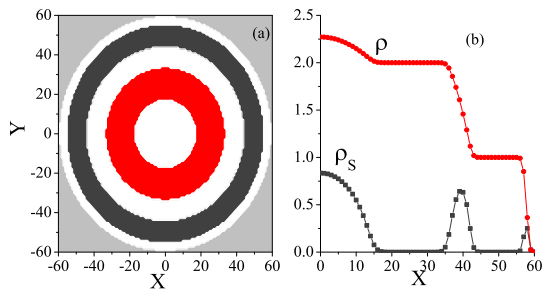,width=8cm,height=5cm}
\caption{(Color online) (a) SF (white) and MI regions [$\rho=2$
(red) and $\rho=1$ (black)] annuli formed in a 2D planar section
${\cal P}_z$ through the 3D lattice, at a vertical distance $z$ from
the center (here $z = 0$) and (b) the corresponding radial variation
of density $\rho_i$ (red circles) and superfluid density $\rho^s_i$
(black squares) for $\mu/E_r=1$, $V_T/E_r=0.0003$ and $V_0=15E_r$;
the outermost gray regions in (a) contain no bosons.}
\label{fig:shells}
\end{figure}

\subsection{Results for the spinless Bose-Hubbard model}

First we compare our mean-field (MF) results with those obtained by
quantum-Monte-Carlo (QMC) simulations in two dimensions~\cite{wessel}. These
simulations use $U/t=25$, $\mu/U=0.37$, and $V_T/U=0.002$, and obtain the
local density $\rho_i$ and the local compressibility
$\kappa^{local}_i=\partial \rho_i/\partial \mu_i$. For this set of parameters
we calculate $\rho_i$ and thence $\kappa^{local}_i$; we also obtain the local
superfluid density $\rho^s_i\equiv \psi^2_i$ (the last formula is valid at
the level of our MF theory). In Fig.~\ref{fig:qmc} we plot versus $R_i$ our
MF results for $\rho_i$, $\kappa^{local}_i$, and $\rho^s_i$ along with data
from QMC simulations~\cite{wessel}. For this set of parameters, the central
region near the origin of the lattice is in the MI phase, i.e., the local
density $\rho_i=1$ and both $\rho^s_i$ and $\kappa^{local}_i$ vanish. This
central core is enveloped by an SF shell, with nonzero values for $\rho^s_i$
and $\kappa^{local}_i$.  As we move radially outward from the center,
$\rho_i$ decreases monotonically till it goes to zero, as do $\rho^s_i$ and
$\kappa^{local}_i$, in the region where $\mu^{eff}_i <0$. The quantitative
agreement between our MF results and those from QMC is shown in
Fig.~\ref{fig:qmc}; there is only a slight discrepancy between the MF
$\rho_i$ and its QMC analog at the MI-SF interface; our result for
$\kappa^{local}_i$ also seems to miss, at this interface, the shoulder that
appears in the QMC $\kappa^{local}_i$ perhaps because our MF theory
overestimates the stability of the SF phase.

\begin{figure}[htbp]
\centering \epsfig{file=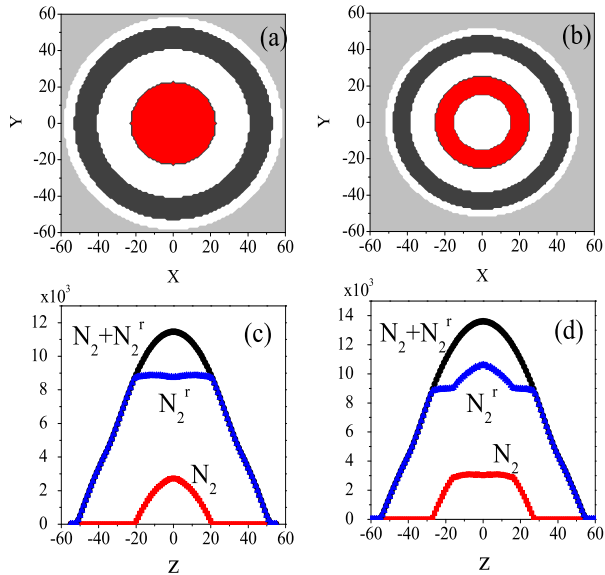,width=9cm,height=9cm}
\caption{(Color online) SF (white) and MI regions [$\rho=2$(red) and
$\rho=1$(black)] annuli formed in the 2D ${\cal P}_{z=0}$ sections
in Figs.\ref{fig:shells1}(a) and (b) for $V_0/E_r=15$,
$V_T/E_r=0.0003$, and (a) $\mu/E_r=0.8$ ($N=7.1\times 10^5$) and (b)
$\mu/E_r=0.9$ ($N=8.9\times 10^5$).  The corresponding integrated
in-trap density profiles $N_2(z)$ (red squares), $N_2^r(z)$ (blue
squares) and $N_2+N_2^r$ (black circles) are shown, respectively, in
(c) and (d); these figures are qualitatively similar to Figs. 3(c)
and (d) in recent experimental study~\cite{bloch}. }
\label{fig:shells1}
\end{figure}

This good agreement between our MF results and those of QMC simulations has
encouraged us to use our MF theory in cases where such simulations pose a
significant numerical challenge.  In particular, we use our theory to make
direct comparisons with experiments~\cite{bloch} that have observed
alternating MI and SF shells in 3D optical lattices by recording in-trap
density distributions of bosons at different filling fractions. We use a
simple-cubic lattice with $121^3$ sites, $\mu/E_r=1$, $V_T/E_r=0.0003$, and
the optical potential $V_0/E_r$ in the range $12-16$ so that the number of
bosons $N$ $\simeq 10^6$, which is comparable to the number of atoms in the
experiments~\cite{bloch} we consider.  This choice of parameters leads to two
well-developed MI shells ($\rho=1$ and $2$, respectively). The MI and SF
shells appear as annuli~\cite{bloch} in a 2D planar section ${\cal P}_z$
through the 3D lattice, at a vertical distance $z$ from the center [see,
e.g., Fig.~\ref{fig:shells}(a) for $V_0/E_r=15$ and $z = 0$ where the core
region is in the SF phase].  Figure \ref{fig:shells}(b) shows that, as we
move radially outward, $\rho_i$ decreases monotonically and $\rho^s_i$ is
zero in the two MI regimes ($16 < R_i < 34$ and $44 < R_i <52$) in which
$\rho_i$ is pinned at $2$ and $1$, respectively.  SF and MI shells alternate
and the outermost one is always in the SF phase; their positions and radii
depend on $\mu$, which also controls the total number $N$ of atoms in the
system, as illustrated by the ${\cal P}_{z=0}$ sections in
Figs.\ref{fig:shells1}(a) and (b) for $V_0/E_r=15$ and $\mu=0.8$
($N=7.1\times 10^5$) and $\mu=0.9$ ($N=8.9\times 10^5$), respectively.  For
any 2D planar section ${\cal P}_z$ we can calculate $N_m(z)$, the number of
bosons in the $\rho=m$ MI annulus, and $N^r_m(z)$, the remaining number of
bosons; the total number of bosons in this planar section is $N(z) = N_m(z) +
N^r_m(z)$, which does not depend on $m$.  In Figs.\ref{fig:shells1} (c) and
(d) we show, for $m=2$ and $\mu=0.8$ and $0.9$, respectively, plots versus
$z$ of $N_m(z)$ (full red squares), $N^r_m(z)$ (full blue triangles), and
their sum $N(z)$ (full black circles). Figures~\ref{fig:shells} (c) and (d)
are remarkably similar to the density profiles obtained in
experiments~\cite{bloch} [cf., their Figs. 3(c) and (d)].

\begin{figure}[htbp]
\centering \epsfig{file=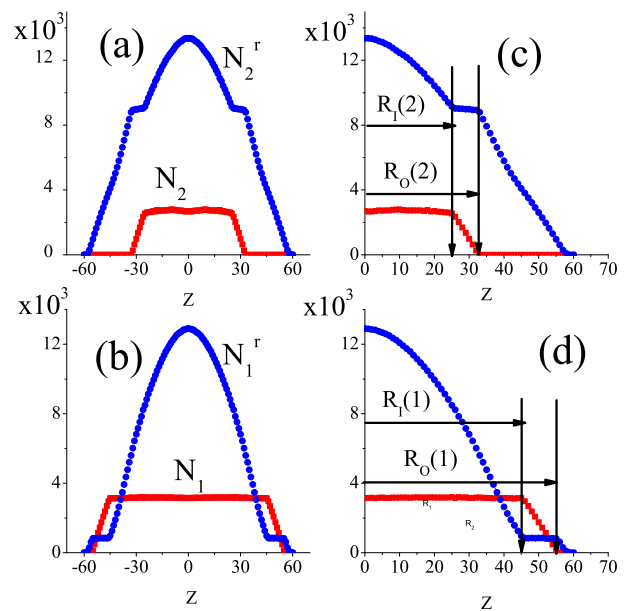,width=9cm,height=9cm}
\caption{(Color online) (a) and (b) Plots of $N_m(z)$(red
squares) and $N_m^r$ (blue circles) versus $z$ for $m=2$ and $1$,
respectively, with $\mu/E_r=1$; the plots in (c) and (d), which
are the same as the right halves of (a) and (b), respectively,
show how we determine the inner and outer radii ($R_I(m)$ and
$R_O(m)$, respectively) from the plateaux in $N_m(z)$ and $N_m^r(z)$. }
\label{fig:shells2}
\end{figure}

The radii of MI shells follow from such in-trap density profiles: In
Figs. \ref{fig:shells2}(a) and (b) we plot $N_m(z)$ and $N_m^r(z)$
versus $z$ for $m=2$ and $m=1$, respectively, with $\mu/E_r=1$. The
curves $N_m(z)$ show nearly flat plateaux for $-R_I(m) \leq z \leq
R_I(m)$; similar plateaux occur in $N^r_m(z)$ for $R_I(m) \leq \mid
z \mid \leq R_O(m)$ [Figs. \ref{fig:shells2}(a), (c) and (b), (d)
for $m=2$ and $m=1$, respectively]. Here $R_I(m)$ and $R_O(m)$ are
the inner and outer radii of the MI shell with integer density $m$.
Elementary geometry can be used to surmise the existence of these
plateaux from the MI-SF shell structure~\cite{rvpspin1} as we show
below.

$N_m(z)$ is $m$ times the total number of sites inside the $\rho=m$ MI
annulus; this number of sites is well approximated by the area $A(z,m)$ of
this annulus. Thus,
\begin{equation}\label{eq:nmz} N_m(z) = m A(z,m) = m \pi
[R^2_O(z,m)- R^2_I(z,m)],
\end{equation}
where $R_O(z,m)$ and $R_I(z,m)$ are, respectively, the outer and
inner radii of the MI annulus with density $\rho=m$, in the 2D
planar section ${\cal P}_z$.  If $z<R_I(m)$, simple geometry yields
$R^2_I(z,m)=R^2_I(m)-z^2$ and $R^2_O(z,m)=R^2_O(m)-z^2$; therefore,
\begin{equation}
N_m(z)=m\pi[R^2_O(m)- R^2_I(m)] , \label{nmz2}
\end{equation}
whence we conclude that $N_m(z)$ is independent of $z$ when $|z| < R_I (m)$;
this result yields the plateaux in the in-trap density profiles shown in
Figs.~\ref{fig:shells2} (a)-(d); if $|z| > R_O(m)$, the 2D planar section has
no MI shell with density $m$, thus, $N_m(z)=0$, which is also apparent in
these figures. For $z < R_I(m)$, the central parts of the 2D planar sections
${\cal P}_{z}$ show SF shells and $N_m^r=N(z)-N_m(z)$ decreases as we
increase $z$. For $R_I(m) < z < R_O(m)$, the central parts of the 2D planar
sections ${\cal P}_{z}$ show MI shells; the number of bosons in such MI
shells is $N_m(z)$ and it is proportional to the area of this central shell,
namely, $m\pi(R^2_O(m)-z^2)$; thus, $N_m(z)$ decreases as we increase $z$
here; however, $N_m^r=N(z)-N_m(z)$ remains independent of $z$, because of the
simple geometrical arguments given above; i.e., we have plateaux in
$N_m^r(z)$ in the region $R_I(m) < z < R_O(m)$. Finally, for $z>R_O(m)$, the
2D planar section ${\cal P}_{z}$ has no MI shell with density
$m$, from which it follows that $N_m(z)=0$.

\begin{figure}[htbp]
\centering \epsfig{file=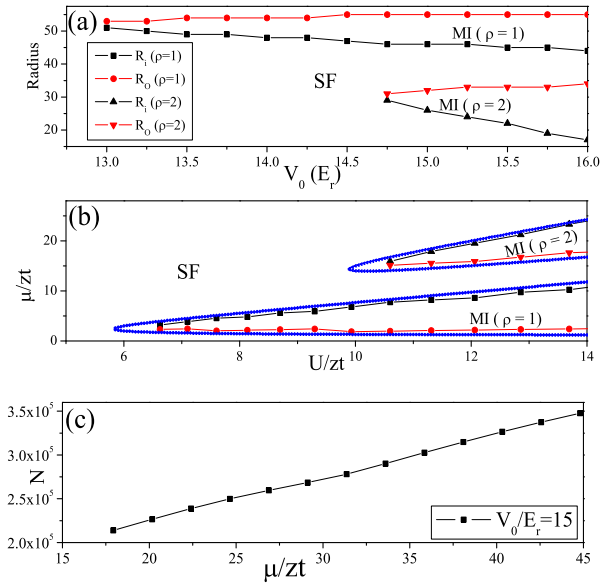,width=9cm,height=9cm}
\caption{(Color online) (a) Plots, for $\rho=m=1$ and $2$, of
$R_O(m)$ and $R_I(m)$ versus $V_0(E_r)$; the MI phase with $\rho=m$
lies between the curves $R_O(m)$ and $R_I(m)$; red circles and red
inverted triangles denote $R_O(1)$ and $R_O(2)$, respectively; and
black squares and black triangles denote $R_I(1)$ and $R_I(2)$,
respectively; (b) Mott-insulating lobes, in plots of
$\mu^+(m)=\mu-V_TR_I^2(m)$ and $\mu^-(m)=\mu-V_T R_O^2(m)$ for $m=2$
and $m=1$ versus $U/(zt)$ (obtained by the conversion
$V_0(E_r)\rightarrow U/zt$); we use the same symbols as in (a); and
we show, for comparison, the boundaries of the MI lobes for $m=1$
(blue diamonds) and $m=2$ (blue triangles) that follow from our
mean-field theory for the \textit{homogeneous} Bose-Hubbard
model~\cite{sheshadri}. (c) An illustrative plot of the total number
of bosons $N$ in the system versus the chemical potential $\mu$.}
\label{fig:radius}
\end{figure}

\begin{figure}[htbp]
\centering \epsfig{file=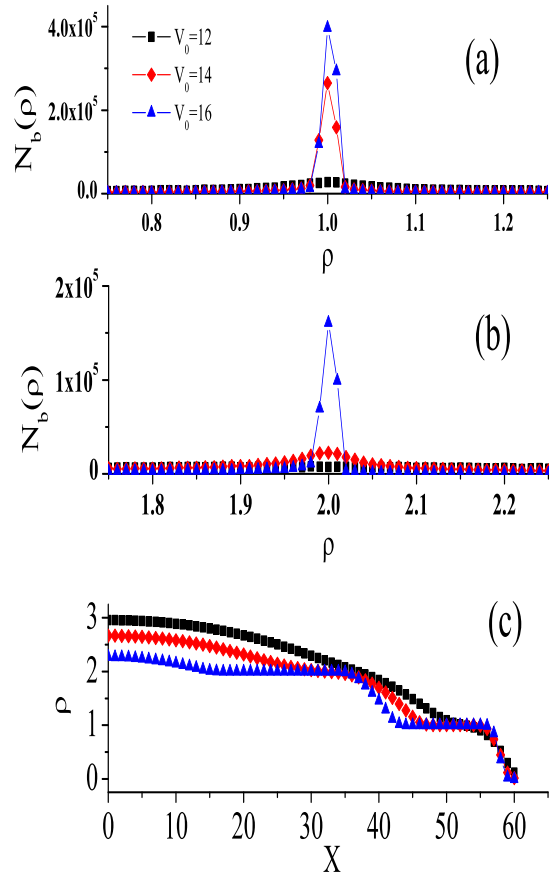,width=8cm,height=13cm} \caption{
(Color online) Representative plots of $N_b(\rho)$, the number of
bosons in the system with a given density $\rho$, versus $\rho$ near
(a) $\rho=1$ and (b) $\rho=2$ for $V_0=12E_r$ (black squares),
$14E_r$ (red diamonds) and $16E_r$ (blue triangles). (c) the radial
variation of density $\rho_i$ for $V_0=12E_r$ (black squares),
$14E_r$ (red diamonds) and $16E_r$ (blue triangles). }
 \label{fig:freq}
\end{figure}

In Fig.~\ref{fig:radius}(a) we plot, for $m=1$ and $2$, $R_O(m)$ and
$R_I(m)$, which we have determined from plots such as those in Figs.
\ref{fig:shells2}(c) and (d), versus $V_0(E_r)$; the MI phase with
$\rho=m$ lies between the curves $R_O(m)$ and $R_I(m)$. Figure
\ref{fig:radius}(a) can be used to obtain the phase diagram of the
homogeneous Bose-Hubbard model as follows: $\mu_i=\mu-V_T R_i^2$, so
$R_O(m)$ and $R_I(m)$ can be used to obtain $\mu^-(m)=\mu-V_T
R_O^2(m)$ and $\mu^+(m)=\mu-V_TR_I^2(m)$, which are, respectively,
the lower and upper boundaries of the Mott lobe with density
$\rho=m$. The resulting Mott lobe (obtained by the conversion
$V_0(E_r)\rightarrow U/zt$) is given in Fig. \ref{fig:radius}(b)
along with its counterpart for the \textit{homogeneous} Bose-Hubbard
model, which we have obtained from the homogeneous mean-field
theory~\cite{sheshadri}; the agreement between these lobes is
striking; and it encourages us to suggest that the phase diagram of
the homogeneous Bose-Hubbard model can be obtained from the inner
and outer radii of the MI shells.  Thus, experiments on cold atoms
in optical lattices {\it with a quadratic confining
potential}~\cite{bloch}, can be used directly to obtain the phase
diagram of the {\it homogeneous} Bose-Hubbard model from $R_O(m)$
and $R_I(m)$, which can be determined for an MI shell with density
$\rho=m$ as described above.  Note that (a) $\mu^-(m)$ and
$\mu^+(m)$ are fixed for a given $V_0(E_r)$ and (b) the total number
of bosons $N$ increases linearly with the chemical potential $\mu$
(see Fig.~\ref{fig:radius}(c)). Therefore, the inner and outer radii
of the MI shell $R_{O,I}(m)=\sqrt{(\mu-\mu^{-,+}/V_T)}$ are
proportional to $\sqrt{N}$, for fixed $V_T$ and $V_0$; this
proportionality has been reported in the recent
experiments~\cite{bloch} [cf., their Fig. 3].

Images of MI shells have been obtained recently from
atomic-clock-shift experiments~\cite{campbell}. By using the
density-dependent transition-frequency shifts, sites with different
densities of bosons can be distinguished spectroscopically; and,
therefore, MI shells, with different values of the integer density
$m$, are revealed as peaks in the occupation number at the
corresponding frequencies. This experiment gives $N_b(\rho)$, the
number of bosons in the system at a given density $\rho$. We use our
inhomogeneous mean-field theory to obtain $N_b(\rho)$ and in Figs.
\ref{fig:freq}(a)-(b) we plot $N_b(\rho)$ (with $\rho$ close to
$m=1$ and $2$, respectively) for $V_0/E_r=12$, $14$ and $16$, with
$V_T/E_r=0.0003$ and $\mu/E_r=1$. The SF and MI shell structure is
evident from the radial variation of the local density given in Fig.
\ref{fig:freq}(c). For $V_0=12E_r$, no Mott shells is developed;
this is reflected in a flat variation of $N_b(\rho)$ for all $\rho$.
However, if $V_0=14E_r$, there is a well-formed MI shell $\rho=1$;
this can be inferred from the peak in $N_b(\rho)$ at $\rho=1$; and,
as $V_0$ increases, more Mott shells, with higher, integral values
of $\rho$, appear. This behavior of $N_b(\rho)$ is in accordance
with recent experiments~\cite{campbell} [cf., their Fig. 1].

\subsection{Results for the Bose-Hubbard model with two species of bosons}

We begin with an investigation of representative phase diagrams of
the Bose-Hubbard model~(\ref{eq:bhab}), with two species of bosons,
in the homogeneous case, i.e., with $V_{Ta}=V_{Tb}=V_T=0$.  These
have been explored to some extent in earlier theoretical
studies~\cite{kuklov03,han04,buonsante08,hu09,ozaki09} and Monte
Carlo simulations~\cite{roscilde07}, but not over as wide a range of
parameters as we consider here.  Next we use the inhomogeneous
mean-field theory that we have developed above to explore
order-parameter profiles and a variety of MI and SF shells that are
obtained when we have a quadratic trap potential. We also present
Fourier transforms of one-dimensional sections of these profiles.

\begin{figure*}[htbp]
\centering \epsfig{file=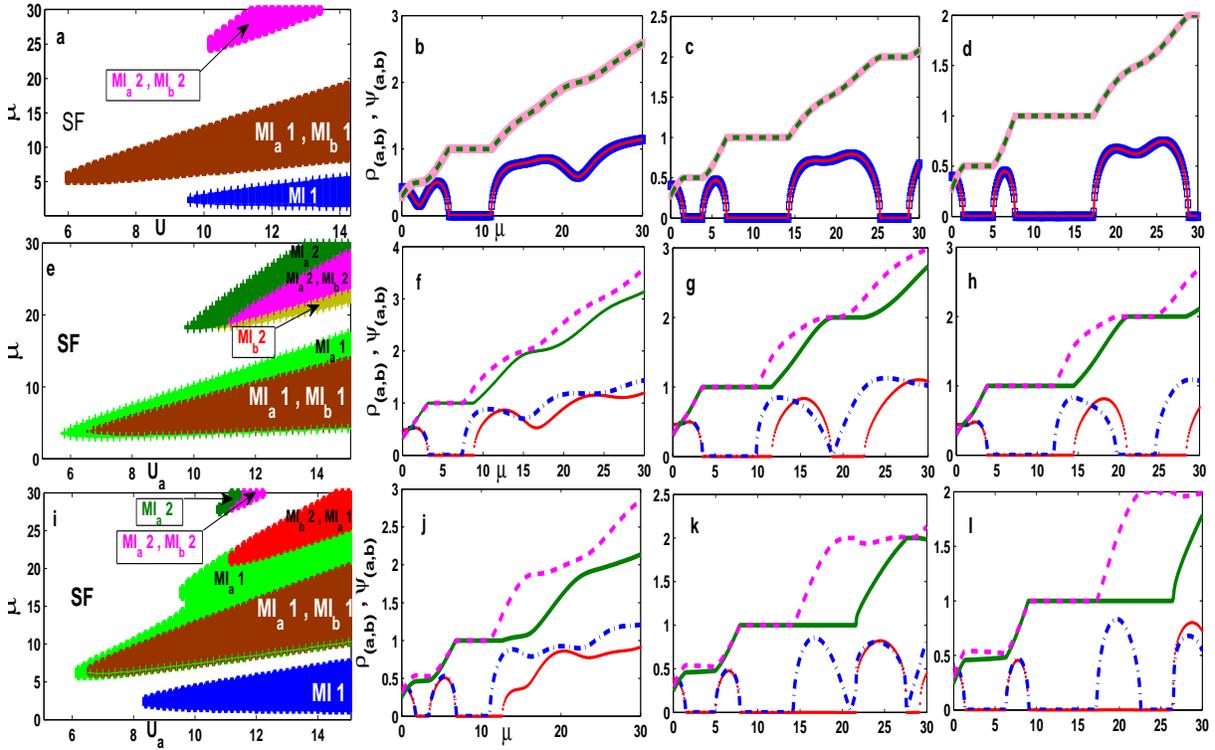,width=16cm,height=10cm}
\caption{(Color online) Phase diagram (a), and plots versus $\mu$ of
the order-parameters $\psi_a$ (red line) and $\psi_b$ (blue dashed
line) and the densities $\rho_a$ (green dashed line) and $\rho_b$
(pink full line) for $U_{ab} = 0.5U_a$, $U_a = U_b$, and $\mu_a =
\mu_b = \mu$ and $U_a = 9$ (b), $U_a = 11$ (c), and $U_a = 13$ (d).
(We do not divide explicitly by $zt$ because we set $zt =1$). The
phase diagram shows an SF phase in which both species are
superfluid; the blue MI$1$ lobe denotes a Mott-insulating phase in
which the density $\rho = 1$ is attained by having $\rho_a = \rho_b
= 1/2$; the brown MI$_a1$MI$_b1$ lobe denotes a Mott-insulating
phase in which the densities $\rho_a = \rho_b = 1$; the pink
MI$_a2$MI$_b2$ lobe denotes a Mott-insulating phase in which the
densities $\rho_a = \rho_b = 2$. In the second row we show the phase
diagram (e), and plots versus $\mu$ of the order-parameters $\psi_a$
and $\psi_b$ and the densities $\rho_a$ and $\rho_b$ for $U_{ab} =
0.2U_a$, $U_b = 0.9U_a$, and $\mu_a = \mu_b = \mu$ and $U_a = 9$
(f), $U_a = 11$ (g), and $U_a = 13$ (h). The phase diagram shows an
SF phase and brown MI$_a1$MI$_b1$ and pink MI$_a2$MI$_b2$ lobes;
these are like their counterparts in (a). In addition we have the
following phases: (i) a green sliver MI$_a1$ in which bosons of type
$a$ are in an MI phase with $\rho_a = 1$ and bosons of type $b$ are
superfluid; (ii) a green-ochre region MI$_b2$ in which bosons of
type $b$ are in an MI phase with $\rho_b = 2$ and bosons of type $a$
are superfluid; and (iii) a dark-green region MI$_a2$ in which
bosons of type $a$ are in an MI phase with $\rho_a = 2$ and bosons
of type $b$ are superfluid. In the third row we show the phase
diagram (i), and plots versus $\mu$ of the order-parameters $\psi_a$
and $\psi_b$ and the densities $\rho_a$ and $\rho_b$ for $U_{ab} =
0.6U_a$, $U_b = 0.9U_a$, and $\mu_a = \mu_b = \mu$ and $U_a = 9$
(j), $U_a = 11$ (k), and $U_a = 13$ (l). The phase diagram shows the
following: an SF phase; blue MI$1$, brown MI$_a1$,MI$_b1$, and pink
MI$_a2$MI$_b2$ lobes; green MI$_a1$ and dark-green MI$_a2$ regions;
these are like their counterparts in (a) and (e). In addition we
have a red MI$_b2$MI$_a1$ lobe in which $\rho_b=2$ and $\rho_a=1$. }
\label{fig:abphases1}
\end{figure*}

First we consider the case $U_{ab} < U_a=U_b$ and $\mu_a = \mu_b =
\mu$ in which the order parameters and densities for both types of
bosons show the same dependence on $\mu$.  In the first row of
Fig.~\ref{fig:abphases1} we show the phase diagram
(Fig.~\ref{fig:abphases1} (a)), and plots versus $\mu$ of the
order-parameters $\psi_a$ (red line) and $\psi_b$ (blue dashed line)
and the densities $\rho_a$ (green dashed line) and $\rho_b$ (pink
full line) for $U_{ab} = 0.5U_a$, $U_a = U_b$, and $\mu_a = \mu_b =
\mu$ and $U_a = 9$ (Fig.~\ref{fig:abphases1} (b)), $U_a = 11$
(Fig.~\ref{fig:abphases1} (c)), and $U_a = 13$
(Fig.~\ref{fig:abphases1} (d)).  (We do not divide explicitly by
$zt$ because we set $zt =1$). The phase diagram shows an SF phase in
which both species are superfluid; the blue MI$1$ lobe denotes a
Mott-insulating phase in which the density $\rho = 1$ is attained by
having $\rho_a = \rho_b = 1/2$; the brown MI$_a1$MI$_b1$ lobe
denotes a Mott-insulating phase in which the densities $\rho_a =
\rho_b = 1$; the    pink MI$_a2$MI$_b2$ lobe denotes a
Mott-insulating phase in which the densities $\rho_a = \rho_b = 2$.
Such phase diagrams can be obtained from plots like those in
Figs.~\ref{fig:abphases1} (b)-(d).

In the second row of Fig.~\ref{fig:abphases1} we show
the phase diagram (Fig.~\ref{fig:abphases1} (e)), and plots
versus $\mu$ of the order-parameters $\psi_a$ and $\psi_b$ and the
densities $\rho_a$ and $\rho_b$ for $U_{ab} = 0.2U_a$,
$U_b = 0.9U_a$, and $\mu_a = \mu_b = \mu$
and $U_a = 9$ (Fig.~\ref{fig:abphases1} (f)),
$U_a = 11$ (Fig.~\ref{fig:abphases1} (g)), and
$U_a = 13$ (Fig.~\ref{fig:abphases1} (h)).
The phase diagram shows an SF phase and brown MI$_a1$MI$_b1$ and
pink MI$_a2$MI$_b2$ lobes; these are like their counterparts
in Fig.~\ref{fig:abphases1} (a). In addition we have the
following phases: (i) a green
sliver MI$_a1$ in which bosons of type $a$ are in an MI phase
with $\rho_a = 1$ and bosons of type $b$ are superfluid;
(ii) a green-ochre region MI$_b2$ in which bosons of type
$b$ are in an MI phase with $\rho_b = 2$ and bosons of type $a$ are
superfluid; and (iii) a dark-green region MI$_a2$ in which bosons of type
$a$ are in an MI phase with $\rho_a = 2$ and bosons of type $b$ are
superfluid.

In the third row of Fig.~\ref{fig:abphases1} we show the phase
diagram (Fig.~\ref{fig:abphases1} (i)), and plots versus $\mu$ of
the order-parameters $\psi_a$ and $\psi_b$ and the densities
$\rho_a$ and $\rho_b$ for $U_{ab} = 0.6U_a$, $U_b = 0.9U_a$, and
$\mu_a = \mu_b = \mu$ and $U_a = 9$ (Fig.~\ref{fig:abphases1} (j)),
$U_a = 11$ (Fig.~\ref{fig:abphases1} (k)), and $U_a = 13$
(Fig.~\ref{fig:abphases1} (l)). The phase diagram shows the
following: an SF phase; blue MI$1$, brown MI$_a1$,MI$_b1$, and pink
MI$_a2$MI$_b2$ lobes; green MI$_a1$ and dark-green MI$_a2$ regions;
these are like their counterparts in Figs.~\ref{fig:abphases1} (a)
and (e). In addition we have a red MI$_b2$MI$_a1$ lobe in which
$\rho_b=2$ and $\rho_a=1$.

\begin{figure*}[htbp]
\centering \epsfig{file=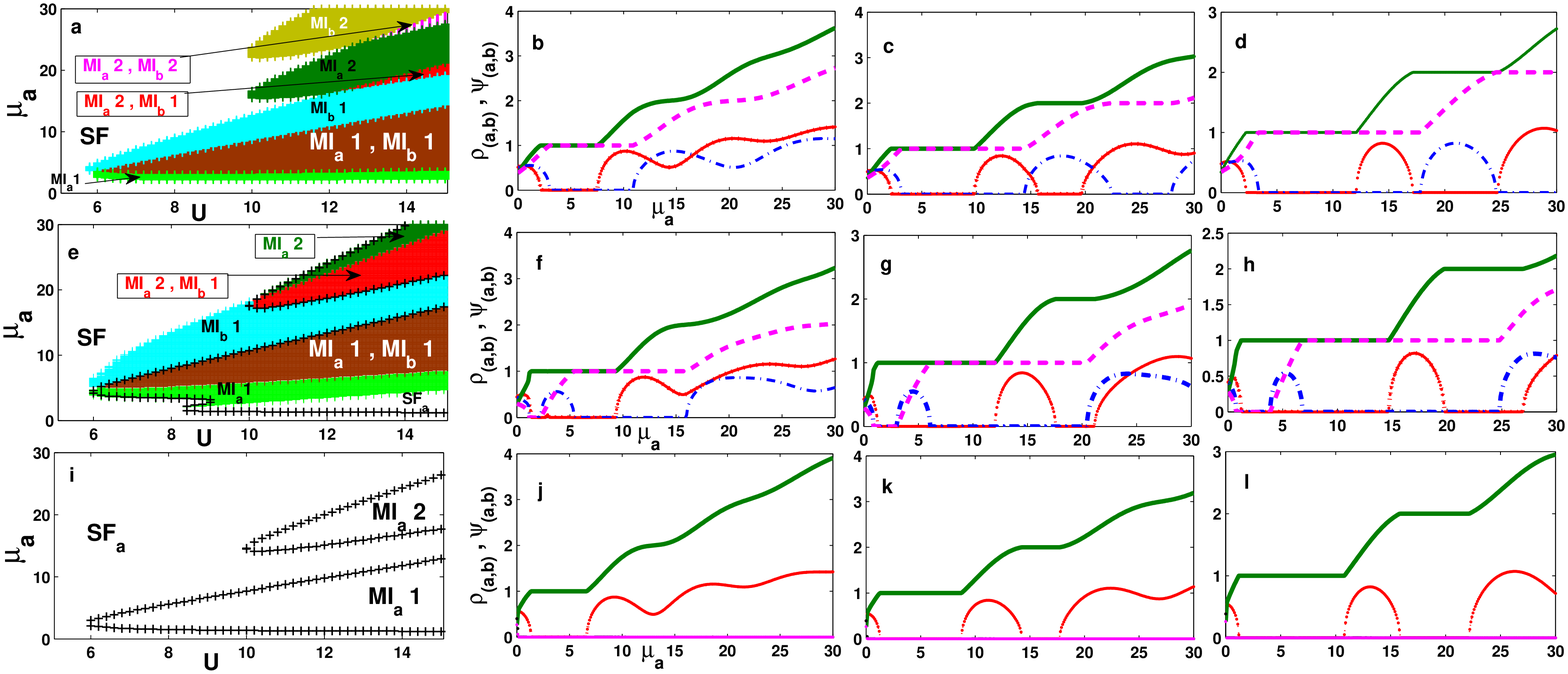,width=16cm,height=10cm}
\caption{(Color online) Phase diagram (a), and plots versus $\mu_a$
of the order-parameters $\psi_a$ (red line) and $\psi_b$ (blue
dashed line) and the densities $\rho_a$ (green full line) and
$\rho_b$ (pink dashed line) for $U_{ab} = 0.1U_a$, $U_a = U_b$, and
$\mu_b = 0.75\mu_a$ and $U_a = 9$ (b), $U_a = 11$ (c), and $U_a =
13$ (d). (We do not divide explicitly by $zt$ because we set $zt
=1$). The phase diagram shows an SF phase and brown MI$_a1$MI$_b1$,
pink MI$_a2$MI$_b2$ and red MI$_a2$MI$_b1$ lobes; green MI$_a1$,
dark-green MI$_a2$ and a green-ochre MI$_b2$ regions; these are like
their counterparts in Figs.~\ref{fig:abphases1} (a) and (e). In
addition we have a light-blue region MI$_b1$ in which bosons of type
$b$ are in an MI phase with $\rho_b = 1$ and bosons of type $a$ are
superfluid. In the second row we show the phase diagram (e), and
plots versus $\mu_a$ of the order-parameters $\psi_a$ and $\psi_b$
and the densities $\rho_a$ and $\rho_b$ for $U_{ab} = 0.3U_a$, $U_b
= U_a$, and $\mu_b = 0.75\mu_a$ and $U_a = 9$ (f), $U_a = 11$ (g),
and $U_a = 13$ (h). The phase diagram shows the following: an SF
phase; brown MI$_a1$,MI$_b1$ and red MI$_a2$MI$_b1$ lobes; green
MI$_a1$, dark-green MI$_a2$ and light-blue MI$_b1$ regions; these
are like their counterparts in (a). In addition we have an SF$_a$
phase in which bosons of type $a$ are in an SF phase and the bosons
density of type $b$ are vanished. In the third row we show the phase
diagram (i), and plots versus $\mu_a$ of the order-parameters
$\psi_a$ and $\psi_b$ and the densities $\rho_a$ and $\rho_b$ for
$U_{ab} = 0.7U_a$, $U_b = U_a$, and $\mu_b = 0.75\mu_a $ and $U_a =
9$ (j), $U_a = 11$ (k), and $U_a = 13$ (l). The phase diagram shows
the following: an SF$_a$ phase; MI$_a1$ and MI$_a2$ regions; these
are like their counterparts in (a) and (e) in which the bosons
density of type $b$ are vanished.} \label{fig:abphases2}
\end{figure*}

Next we consider the case $U_{ab} < U_a$, $U_b = U_a$, and $\mu_b =
0.75\mu_a$. Specifically, in the first row of
Fig.~\ref{fig:abphases2}  we show the phase diagram
(Fig.~\ref{fig:abphases2} (a)), and plots versus $\mu_a$ of the
order-parameters $\psi_a$ (red line) and $\psi_b$ (blue dashed line)
and the densities $\rho_a$ (green full line) and $\rho_b$ (pink
dashed line) for $U_{ab} = 0.1U_a$, $U_a = U_b$, and $\mu_b =
0.75\mu_a$ and $U_a = 9$ (Fig.~\ref{fig:abphases2} (b)), $U_a = 11$
(Fig.~\ref{fig:abphases2} (c)), and $U_a = 13$
(Fig.~\ref{fig:abphases2} (d)).  (We do not divide explicitly by
$zt$ because we set $zt =1$). The phase diagram shows an SF phase
and brown MI$_a1$MI$_b1$, pink MI$_a2$MI$_b2$ and red MI$_a2$MI$_b1$
lobes; green MI$_a1$, dark-green MI$_a2$ and a green-ochre MI$_b2$
regions; these are like their counterparts in
Figs.~\ref{fig:abphases1} (a) and (e). In addition we have a
light-blue region MI$_b1$ in which bosons of type $b$ are in an MI
phase with $\rho_b = 1$ and bosons of type $a$ are superfluid.

In the second row of Fig.~\ref{fig:abphases2} we show the phase diagram
(Fig.~\ref{fig:abphases2} (e)), and plots versus $\mu_a$ of the
order-parameters $\psi_a$ and $\psi_b$ and the densities $\rho_a$ and
$\rho_b$ for $U_{ab} = 0.3U_a$, $U_b = U_a$, and $\mu_b = 0.75\mu_a$ and $U_a
= 9$ (Fig.~\ref{fig:abphases2} (f)), $U_a = 11$ (Fig.~\ref{fig:abphases2}
(g)), and $U_a = 13$ (Fig.~\ref{fig:abphases2} (h)). This phase diagram shows
the following: an SF phase; brown MI$_a1$,MI$_b1$ and red MI$_a2$MI$_b1$
lobes; green MI$_a1$, dark-green MI$_a2$ and light-blue MI$_b1$ regions;
these are like their counterparts in Figs.~\ref{fig:abphases2} (a).  In
addition we have a dark-gray SF$_a$ phase in which bosons of type $a$ are in
an SF phase and the bosons of type $b$ have vanished.

In the third row of Fig.~\ref{fig:abphases2} we show the phase
diagram (Fig.~\ref{fig:abphases2} (i)), and plots versus $\mu_a$ of
the order-parameters $\psi_a$ and $\psi_b$ and the densities
$\rho_a$ and $\rho_b$ for $U_{ab} = 0.7U_a$, $U_b = U_a$, and $\mu_b
= 0.75\mu_a $ and $U_a = 9$ (Fig.~\ref{fig:abphases2} (j)), $U_a =
11$ (Fig.~\ref{fig:abphases2} (k)), and $U_a = 13$
(Fig.~\ref{fig:abphases2} (l)).  The phase diagram shows the
following: an SF$_a$ phase; MI$_a1$ and MI$_a2$ regions; these are
like their counterparts in Figs.~\ref{fig:abphases2} (a) and (e) in
which the bosons density for type $b$ has vanished.

\begin{figure*}[htbp]
\centering \epsfig{file=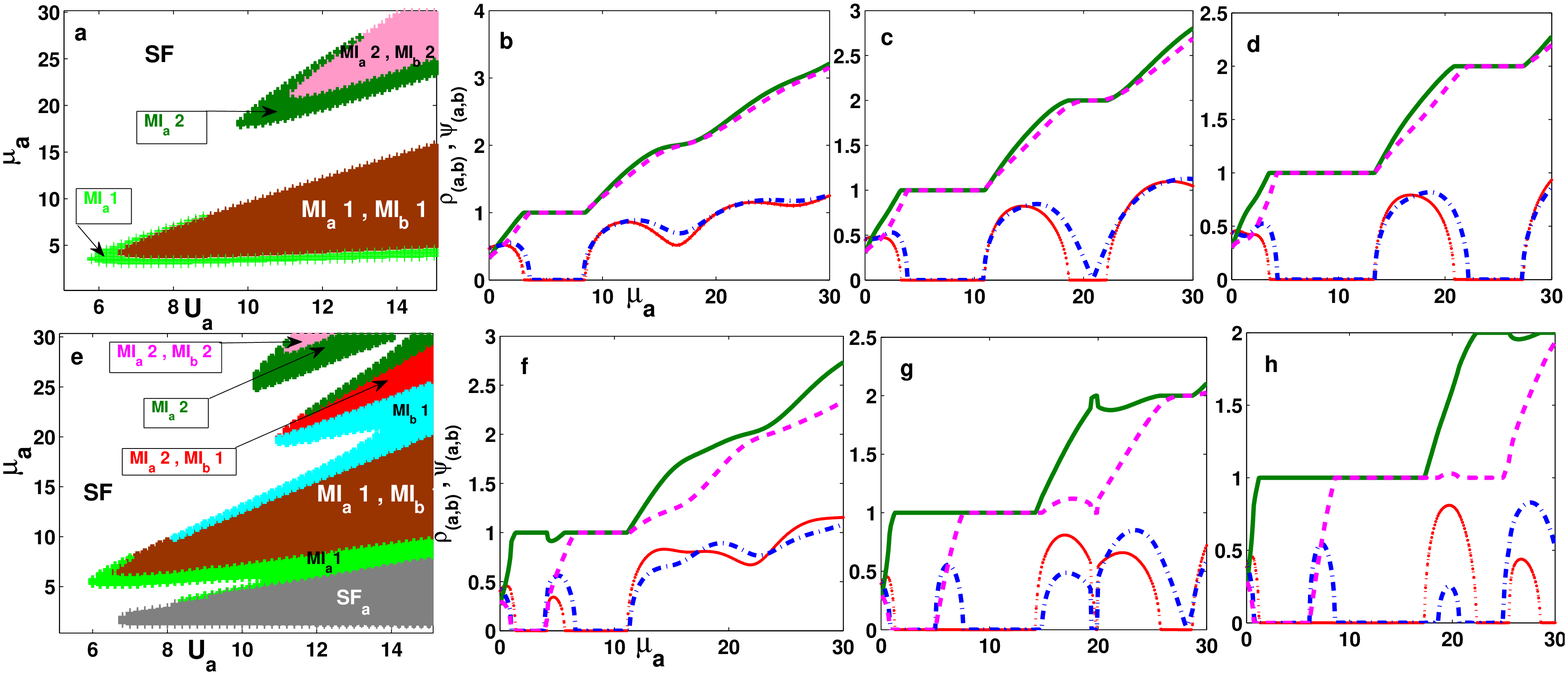,width=16cm,height=10cm}
\caption{(Color online) Plots versus $\mu_a$ of the order-parameters
$\psi_a$ and $\psi_b$ and the densities $\rho_a$ and $\rho_b$ for
$U_{ab} = 0.5U_a$, $U_b = 0.9U_a$, and $\mu_b = 0.9\mu_a$ and $U_a =
9$ (f), $U_a = 11$ (g), and $U_a = 13$ (h).  The phase diagram shows
the following: an SF phase and SF$_a$; brown MI$_a1$,MI$_b1$, red
MI$_a2$MI$_b1$ and pink MI$_a2$MI$_b2$ lobes; green MI$_a1$,
dark-green MI$_a2$ and light-blue MI$_b1$ regions; these are like
their counterparts in Figs.~\ref{fig:abphases2} (a) and (e).}
\label{fig:abphases3}
\end{figure*}
\begin{figure*}[htbp]
\centering \epsfig{file=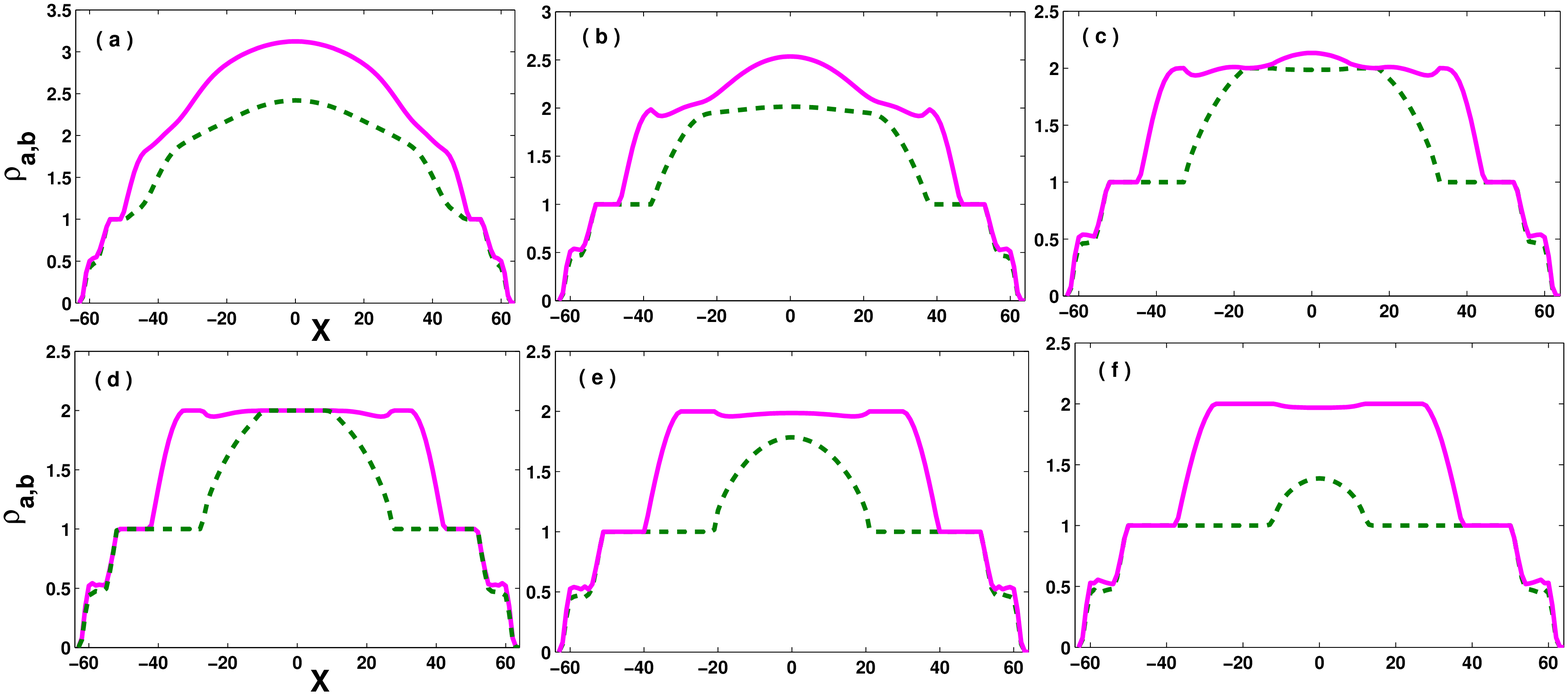,width=16cm,height=10cm}
\caption{(Color online) Plots of $\rho_{ai}$ (green dashed line) and
$\rho_{bi}$ (pink line) versus $X$, along the $Y=Z=0$ line,  with
$\mu_a =\mu_b = 30$, $V_T/(zt)=0.008$, and (a) $U_a/(zt)=8$ , $U_b
=0.9U_a$ , and $U_{ab} = 0.6U_a$ (b) $U_a/(zt)=10$ , $U_b =0.9U_a$ ,
and $U_{ab} = 0.6U_a$, (c) $U_a/(zt)=11$ , $U_b =0.9U_a$ , and
$U_{ab} = 0.6U_a$, (d)$U_a/(zt)=12$ , $U_b =0.9U_a$ , and $U_{ab} =
0.6U_a$, (e) $U_a/(zt)=13$ , $U_b =0.9U_a$ , and $U_{ab} = 0.6U_a$,
and (f) $U_a/(zt)=14$ , $U_b =0.9U_a$ , and $U_{ab} = 0.6U_a$.}
\label{fig:fig2b5}
\end{figure*}
\begin{figure*}[htbp]
\centering \epsfig{file=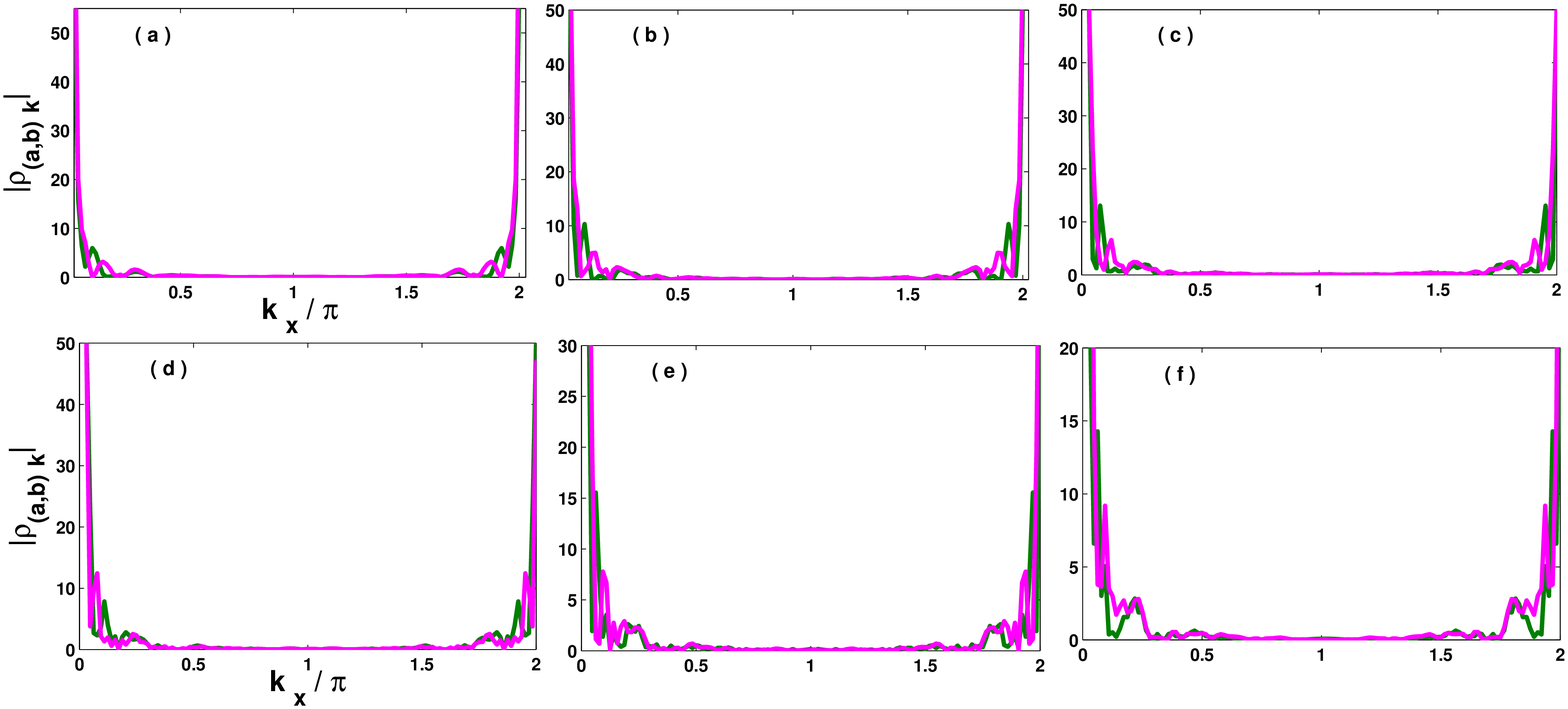,width=16cm,height=10cm}
\caption{(Color online) The moduli of the one-dimensional Fourier
transforms, namely, $|\rho_{(a,b)k}|$, of the plots of
$\rho_{(a,b)i}$ in Figs.~\ref{fig:fig2b5}(a)- (f) are plotted,
respectively, in (a) - (f) here versus the wave vector $k_X/\pi$.}
\label{fig:fig2b6}
\end{figure*}

\begin{figure*}[htbp]
\centering \epsfig{file=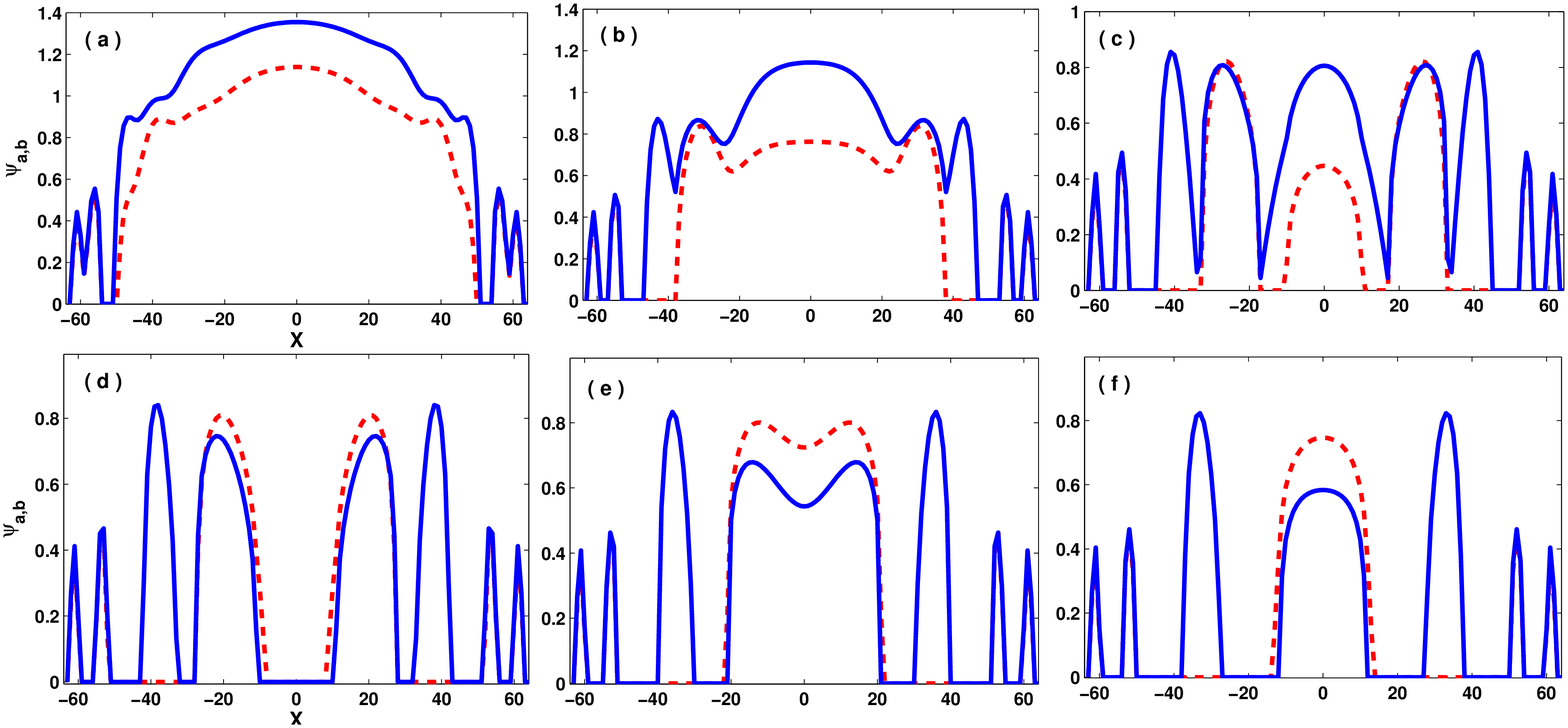,width=16cm,height=10cm}
\caption{(Color online) Plots of Plots of $\psi_{ai}$ (green dashed
line) and $\psi_{bi}$ (pink line) versus $X$, along the $Y=Z=0$
line, with $\mu_a =\mu_b = 30$, $V_T/(zt)=0.008$, and (a)
$U_a/(zt)=8$ , $U_b =0.9U_a$ , and $U_{ab} = 0.6U_a$ (b)
$U_a/(zt)=10$ , $U_b =0.9U_a$ , and $U_{ab} = 0.6U_a$, (c)
$U_a/(zt)=11$ , $U_b =0.9U_a$ , and $U_{ab} = 0.6U_a$,
(d)$U_a/(zt)=12$ , $U_b =0.9U_a$ , and $U_{ab} = 0.6U_a$,  (e)
$U_a/(zt)=13$ , $U_b =0.9U_a$ , and $U_{ab} = 0.6U_a$, and (f)
$U_a/(zt)=14$ , $U_b =0.9U_a$ , and $U_{ab} = 0.6U_a$. }
\label{fig:fig2b7}
\end{figure*}
\begin{figure*}[htbp]
\centering \epsfig{file=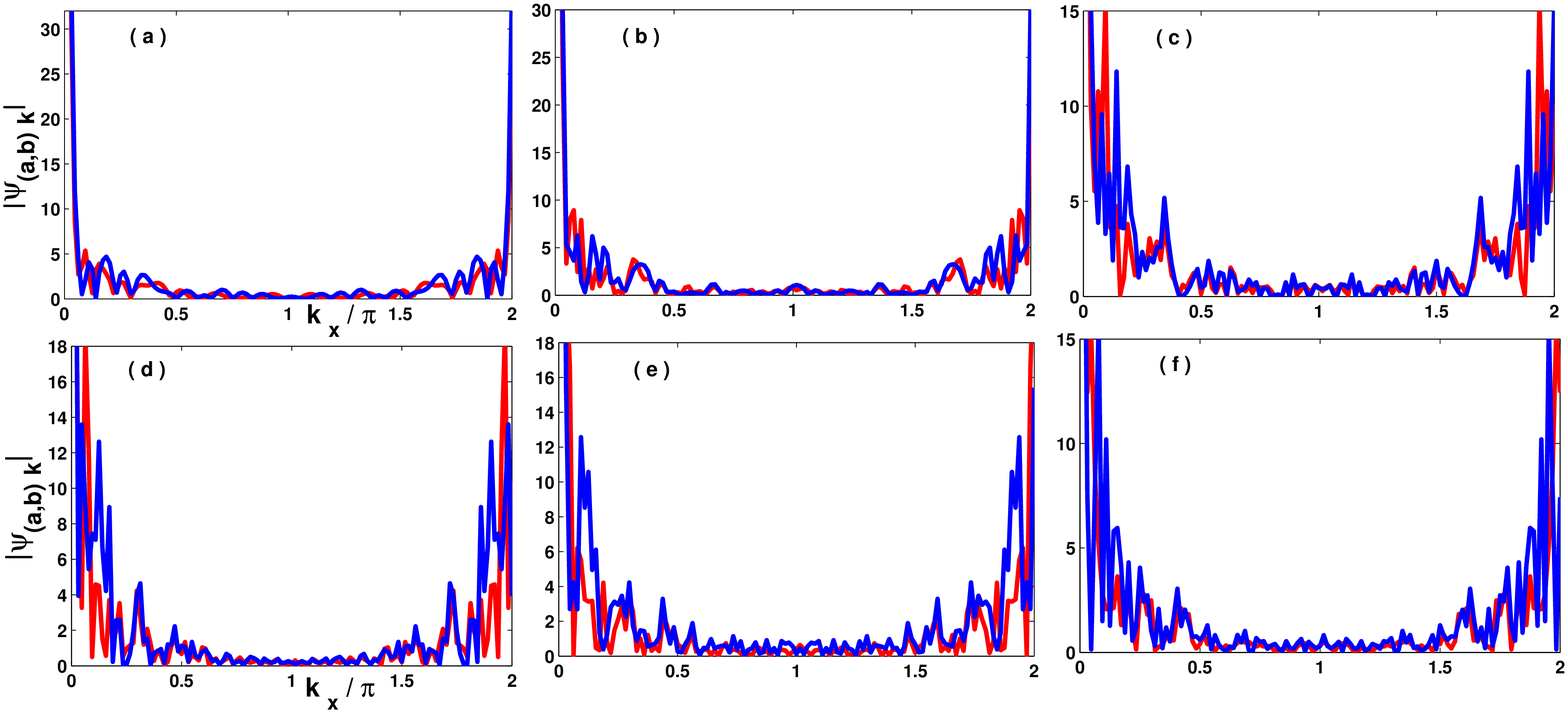,width=16cm,height=10cm}
\caption{(Color online) The moduli of the one-dimensional Fourier
transforms, namely, $|\psi_{(a,b)k}|$, of the plots of
$\psi_{(a,b)i}$ in Figs.~\ref{fig:fig2b7}(a) - (f) are plotted,
respectively, in (a) - (f) here versus the wave vector $k_X/\pi$.}
\label{fig:fig2b8}
\end{figure*}

\begin{figure*}[htbp]
\centering \epsfig{file=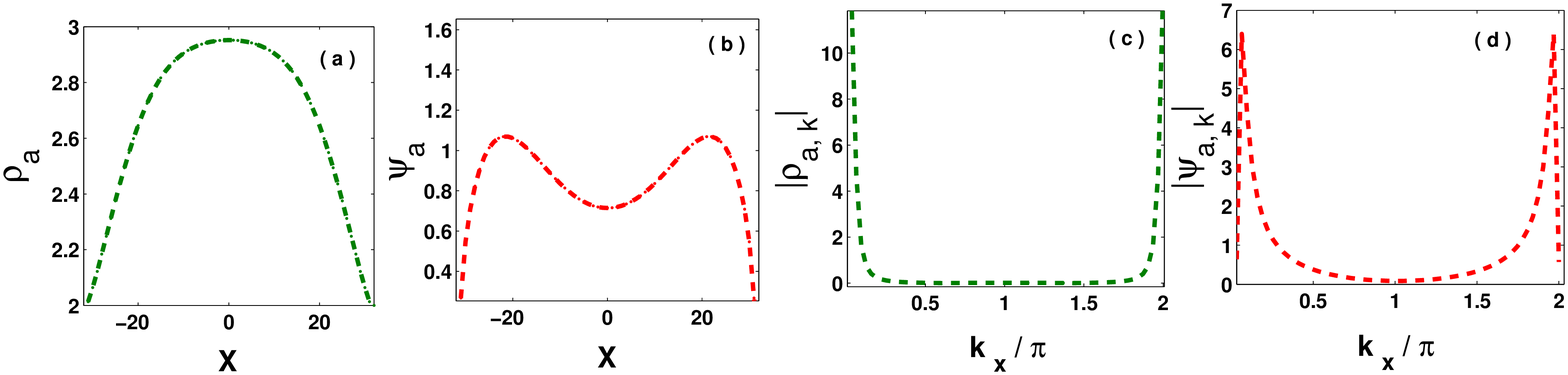,width=14cm,height=8cm}
\caption{(Color online) Plots of (a) $\rho_{a,i}$ (b) $\psi_{a,i}$
versus $X$, along the $Y=Z=0$ line,  with $U_{ab}=2.22U_a,\,
U_b=0.65U_a,\,  U_a=13,\, \mu_b=0.8\mu_a, V_T=0.008$, and $L=64$
(with this set of parameter values $\rho_b$ and $\psi_b$ vanish);
corresponding plots of the moduli of the one-dimensional Fourier
transforms, namely, (c) $|\rho_{a,k}|$, and (d) $|\psi_{a,k}|$
versus the wave vector $k_X$.} \label{fig:uablarge}
\end{figure*}

\begin{figure*}[htbp]
\centering \epsfig{file=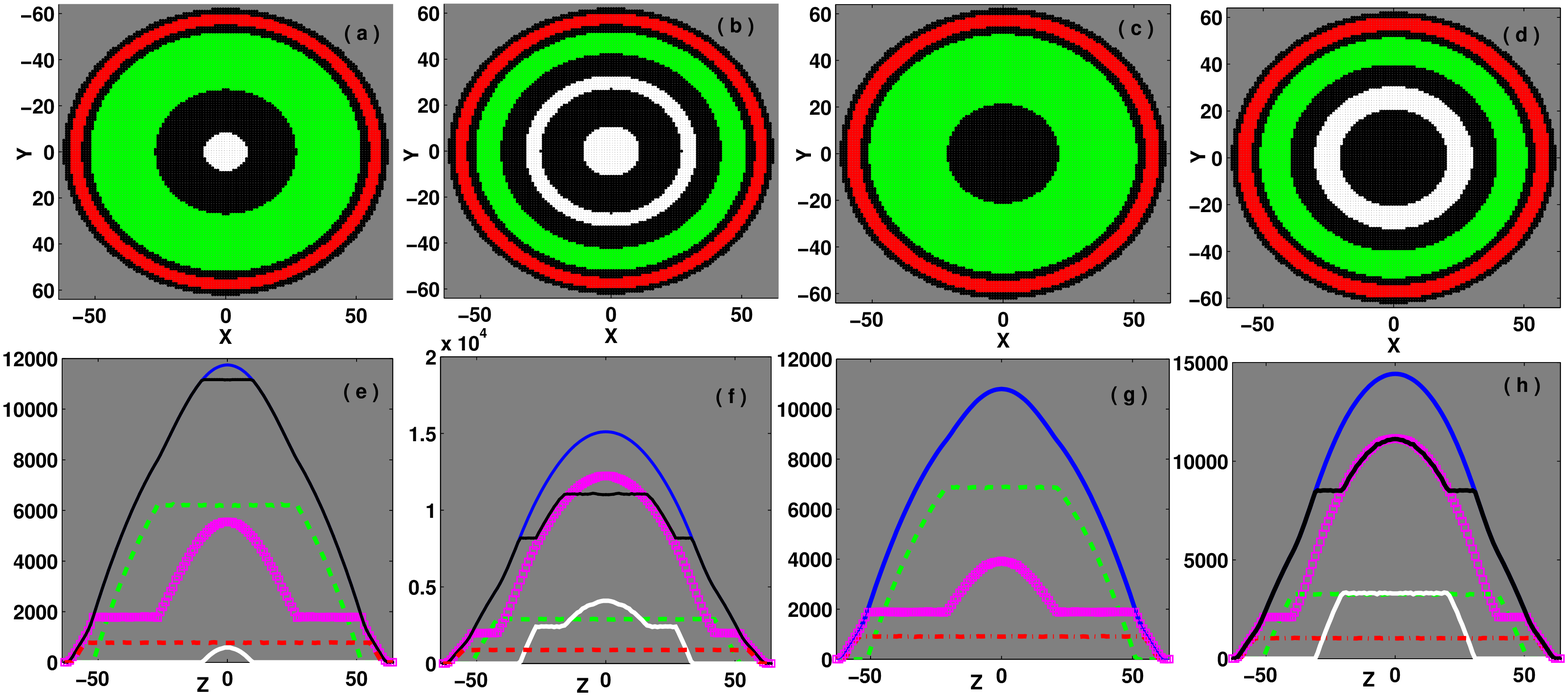,width=16cm,height=9cm}
\caption{(Color online) SF (black), MI2 (white), MI1 (green),
MI(a,b) 1/2 (red) annuli in the 2D planar section  ${\cal P}_{Z=0}$
(see text) with $U_{a}/(zt) = 13$, $U_{b}=0.9U_{a}$,
$U_{ab}=0.6U_{a}$ , $V_T/(zt)=0.008$ and $\mu_a=\mu_b= 30$ (a) type
a (b) type b; for these parameter values, the  integrated, in-trap
density profiles are plotted versus $Z$ in (e) and (f),
respectively. These in-trap profiles show the total number of bosons
for type $a$ and $b$; $N_{{a}_T}$ ,$N_{{b}_T}$ (blue full lines),
the number of bosons in MI2 and MI1 regions, $N_{{a}_2}$,
$N_{{b}_2}$ (white line in (e) and (f)) and $N_{{a}_1}$ ,
$N_{{b}_1}$(green dash line in (e) and (f)), respectively, the
numbers of bosons in MI(a,b) 1/2 (red dash line in (e) and (f))
regions, [$N_{{a}_T}-N_{{a}_2}$, $N_{{b}_T}-N_{{b}_2}$] (black full
line in (e) and (f)), and [$N_{{a}_T}-N_{{a}_1}$
,$N_{{b}_T}-N_{{b}_1}$] (pink annulus  in (e) and (f)).   The
annular structures for the profiles of Figs.~\ref{fig:fig2b5} (e)
and ~\ref{fig:fig2b7} (e) are given in (c) and (d) and corresponding
integrated, in-trap density profiles in (g) and (h), respectively.
The outermost gray regions in (a) - (d) contain no bosons.}
\label{fig:fig2b9}
\end{figure*}

We now consider the case $U_{ab} < U_a$, $U_b = 0.9U_a$, and $\mu_b
= 0.9\mu_a$. Specifically, in the first row of
Fig.~\ref{fig:abphases3} we show the phase diagram
(Fig.~\ref{fig:abphases3} (a)), and plots versus $\mu_a$ of the
order-parameters $\psi_a$ (red line) and $\psi_b$ (blue dashed line)
and the densities $\rho_a$ (green full line) and $\rho_b$ (pink
dashed line) for $U_{ab} = 0.2U_a$, $U_a = 0.9U_b$, and $\mu_b =
0.9\mu_a$ and $U_a = 9$ (Fig.~\ref{fig:abphases3} (b)), $U_a = 11$
(Fig.~\ref{fig:abphases3} ( c)), and $U_a = 13$
(Fig.~\ref{fig:abphases3} (d)).  The phase diagram shows an SF phase
and brown MI$_a1$MI$_b1$ and pink MI$_a2$MI$_b2$ lobes; green
MI$_a1$ and dark-green MI$_a2$; these are like their counterparts in
Figs.~\ref{fig:abphases1} and Figs.~\ref{fig:abphases2} (a) and (e).

In the second row of Fig.~\ref{fig:abphases3} we show the phase
diagram (Fig.~\ref{fig:abphases3} (e)), and plots versus $\mu_a$ of
the order-parameters $\psi_a$ and $\psi_b$ and the densities
$\rho_a$ and $\rho_b$ for $U_{ab} = 0.5U_a$, $U_b = 0.9U_a$, and
$\mu_b = 0.9\mu_a$ and $U_a = 9$ (Fig.~\ref{fig:abphases3} (f)),
$U_a = 11$ (Fig.~\ref{fig:abphases3} (g)), and $U_a = 13$
(Fig.~\ref{fig:abphases3} (h)).  The phase diagram shows the
following: an SF phase and SF$_a$; brown MI$_a1$,MI$_b1$, red
MI$_a2$MI$_b1$ and pink MI$_a2$MI$_b2$ lobes; green MI$_a1$,
dark-green MI$_a2$ and light-blue MI$_b1$ regions; these are like
their counterparts in Figs.~\ref{fig:abphases2} (a) and (e).

We now consider the effect of a parabolic potential and use the
inhomogeneous mean-field theory, developed in the previous Section,
to obtain alternating spherical shells of the variety of MI and SF
phases, shown in the phase diagrams in Figs.~\ref{fig:abphases1},
\ref{fig:abphases2}, and \ref{fig:abphases3}, for the two-species BH
model~(\ref{eq:bhab}). We do this by obtaining the order-parameter
profiles $\{\psi_{ai},\rho_{ai} ;\psi_{bi},\rho_{bi}\}$ and also by
obtaining in-trap density distributions of bosons at representative
values of $U_{ab}, \, U_{a}, \, U_{b}$, and $\mu_{a} = \mu_{b}$. In
particular, we use a 3D simple-cubic lattice with $128^3$ sites and
$V_T/(zt)=0.008$; and we study the following representative case:
$\mu_{a}/(zt)=\mu_{b}/(zt)= 30$ , $U_{ab}=0.6U_{a}$ ,
$U_{b}=0.9U_{a}$, when $U_{a}/(zt)=13$. With these parameters the
total number of bosons $N_T\simeq 10^6$, which is comparable to
experimental values.  Furthermore, this choice of parameters leads
not only to SF shells but also two well-developed MI shells (MI1 and
MI2) .

We show representative plots of the densities $\rho_{a}$ (green
dashed line) and $\rho_{b}$ (pink line) versus the position $X$
along the line $Y=Z=0$ are shown in Figs.~\ref{fig:fig2b5} for
$\mu_a =\mu_b = 30$, $V_T/(zt)=0.008$, and the following six
parameter sets, respectively: (a) $U_a/(zt)=8$, $U_b =0.9U_a$, and
$U_{ab} = 0.6U_a$ (b) $U_a/(zt)=10$, $U_b =0.9U_a$, and $U_{ab}=
0.6U_a$, (c) $U_a/(zt)=11$, $U_b =0.9U_a$, and $U_{ab} = 0.6U_a$,
(d) $U_a/(zt)=12$, $U_b =0.9U_a$, and $U_{ab} = 0.6U_a$,  (e)
$U_a/(zt)=13$, $U_b =0.9U_a$, and $U_{ab} = 0.6U_a$, and (f)
$U_a/(zt)=14$, $U_b =0.9U_a$, and $U_{ab} = 0.6U_a$.  It is also
useful to obtain a complementary, Fourier-representation picture of
the profiles in Figs.~\ref{fig:fig2b5} (a)-(f) because it might be
possible to obtain them in time-of-flight measurements~\cite{rmp}.
Three-dimensional transforms of the shell structure can be obtained,
but they are not easy to visualize; therefore, we present the
one-dimensional Fourier transforms of $\rho_a(X,Y=0,Z=0)$ and
$\rho_b(X,Y=0,Z=0)$ with respect to $X$. The moduli of these
transforms, namely, $|\rho_{a,k_X}|$ (green lines), and
$|\rho_{b,k_X}|$ (pink lines) of the profiles in
Figs.~\ref{fig:fig2b5} (a)-(f) are plotted, respectively, in
Figs.~\ref{fig:fig2b6} (a)-(f) versus the wave vector $k_X/\pi$. The
principal peaks in these transforms occur at $k_X = 0$ (or $2 \pi$);
these are associated with the spatially uniform MI and SF phases .
In an infinite system with no confining potential, these are the
only peaks; however, the quadratic confining potential leads to
shells of MI and SF phases (see below); this shell structure leads
to the subsidiary peaks that appear in Figs.~\ref{fig:fig2b6} (a) -
(f) away from $k_X = 0, \ $ and $2 \pi$.

We can also obtain order-parameter-profile plots; these are shown in
Figs.~\ref{fig:fig2b7}(a) - (f) for the same parameter values as in
Figs.~\ref{fig:fig2b5}(a) -(f), respectively; in these plots
$\psi_{a}$ is indicated by a red dashed line and $\psi_{b}$ by a
blue line. We can also obtain the one-dimensional Fourier transforms
of $\psi_a(X,Y=0,Z=0)$ and $\psi_b(X,Y=0,Z=0)$ with respect to $X$.
The moduli of these transforms, namely, $|\psi_{a,k_X}|$ (red lines)
and $|\psi_{b,k_X}|$ (blue lines) of the profiles
Figs.~\ref{fig:fig2b7}(a) - (f) are plotted, respectively, in
Figs.~\ref{fig:fig2b8}(a) - (f) versus the wave vector $k_X/\pi$ .
Again, the principal peaks in these transforms occur at $k_X = 0$
(or $2 \pi$); but subsidiary peaks occur because of the shell
structure imposed by the confining potential.

In Fig.~\ref{fig:uablarge}(a) - (d) we show representative density
and order-parameter profiles, for $\rho_a$ and $\psi_a$, and the
moduli of their Fourier transforms for $U_{ab}=2.22U_a,\,
U_b=0.65U_a,\,  U_a=13,\, \mu_b=0.8\mu_a, V_T=0.008$, and $L=64$.
With this set of parameter values $\rho_b$ and $\psi_b$ vanish.

The density and order-parameter profiles of Figs.~\ref{fig:fig2b5}
and Figs.~\ref{fig:fig2b7} lead to the shell structure that we
describe below. For specificity, consider Fig.~\ref{fig:fig2b5}(d)
and Fig.~\ref{fig:fig2b7}(d). These plots show that along the line
$Y=Z=0$, in the region from $X=0$ to $|X|\equiv 10$ the system has
an MI phase for both types of bosons with $\rho_{b} = \rho_{a} =2$
and $\psi_{a} =\psi_{b} =0$; in the regions $10<X<28$ and
$-28<X<-10$), the system displays an SF phase for both types of
bosons with $\rho_{b} > \rho_{a}$ and slightly $\psi_{a}
>\psi_{b} > 0$;  in the intervals $28<X<50$ and $-50<X<-28$ bosons of type
${a}$ are in an MI phase with $\rho_{a}=1$; when $28<X<34$ or
$-34<X<-28$ bosons of type $b$ are in an MI phase with $\rho_{b}=2$;
in the regions $34<X<42$ and $-42<X<-34$ bosons of type $b$ are in
an SF phase whereas those of type $a$ are still in the MI phase with
$\rho_{a}=1$; if $42<X<50$ or $-50<X<-42$, then the displays MI
phase for both types of bosons with $\rho_{b} = \rho_{a} =1$; at
slightly larger values of $|X|$ the system moves into an SF phase
with $\rho_{a} = \rho_{b} > 0$ and $\psi_{a} = \psi_{b} >0$; at even
larger values of $|X|$ the system moves into a very narrow MI phase
for the both types of bosons with $\rho_{a} = \rho_{b} =0.5$, such
that the total density for the system is $\rho =\rho_{a} + \rho_{b}
=1$  and $\psi_{a} = \psi_{b} = 0$; as $|X|$ increases further, the
system displays a very narrow SF region until it enters a small
region in which the boson density vanishes for the both types of
bosons.

From the profiles in Figs.~\ref{fig:fig2b5} and
Figs.~\ref{fig:fig2b7} (a) - (f) it is clear that the precise
sequence and types of MI and SF shells depends on the parameters in
the BH model~(\ref{eq:bhab}) for two species of bosons.  These SF
and MI shells appear as annuli in a two-dimensional (2D) planar
section ${\cal P}_Z$ through the 3D lattice, at a vertical distance
$Z$ from the center as shown, for ${\cal P}_{Z=0}$, in
Figs.~\ref{fig:fig2b9} (a) and (b), for bosons of types $a$ and $b$,
respectively for the profiles of Figs.~\ref{fig:fig2b5}(d) and
~\ref{fig:fig2b7}(d). Similar annular structures for the profiles of
Figs.~\ref{fig:fig2b5} (e) and ~\ref{fig:fig2b7} (e) are given in
Figs.~\ref{fig:fig2b9} (c) and (d). We do not display the annular
structures for the other profiles in Figs.~\ref{fig:fig2b5} and
Figs.~\ref{fig:fig2b7}.

For any 2D planar section ${\cal P}_Z$ we can calculate integrated,
in-trap density profiles such as $N_m(Z)$, the number of bosons in
the $\rho=m$ MI annuli, as we discussed for the BH model with one
species of bosons. Here $m$ is an integer; we concentrate on $m=1$
or $m=2$. We can also calculate the remaining number of bosons,
e.g., $N^r_{am}(Z)=N_{aT}-N_{am}(Z)$ or
$N^r_{bm}(Z)=N_{bT}-N_{bm}(Z)$. For the parameter values of
Figs.~\ref{fig:fig2b9} (a) and (b), illustrative integrated, in-trap
density profiles are plotted versus $Z$ in Figs.~\ref{fig:fig2b9}
(e) and (f), respectively. These in-trap profiles show the total
number of bosons for type $a$ and $b$, $N_{aT}$ and $N_{bT}$(blue
full lines), the number of bosons in MI2 and MI1 regions,
$N_{{a}_2}$, $N_{{b}_2}$ (white line in (e) and (f)) and $N_{{a}_1}$
, $N_{{b}_1}$(green dash line in (e) and (f)), respectively, the
numbers of bosons in $MI_{(a,b)}$ 1/2 (red dash line in (e) and (f))
regions, [$N_{{a}_T}-N_{{a}_2}$, $N_{{b}_T}-N_{{b}_2}$] (black full
line in (e) and (f)), and [$N_{{a}_T}-N_{{a}_1}$
,$N_{{b}_T}-N_{{b}_1}$] (pink annulus)  in (e) and (f).  The
outermost gray regions contain no bosons. Integrated, in-trap
density profiles for the planar sections in Figs.~\ref{fig:fig2b9}
(c) and (d) are shown in Figs.~\ref{fig:fig2b9} (g) and (h),
respectively.

\subsection{Results for the Spin-1 Bose-Hubbard Model}

\begin{figure}[htbp]
\centering \epsfig{file=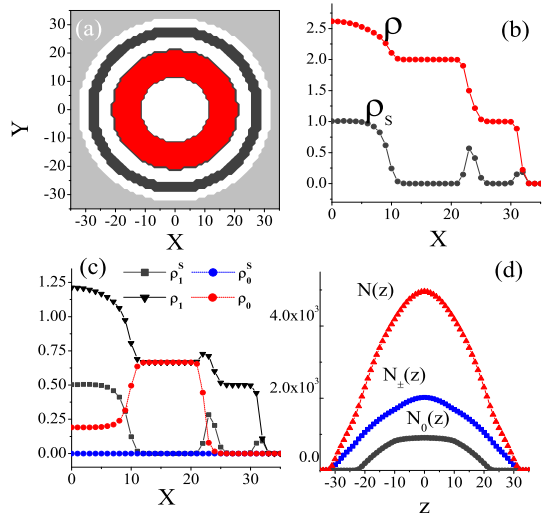,width=8cm,height=8cm}
\caption{(Color online) (a) SF (white) and MI regions [$\rho=2$(red)
and $\rho=1$(black)] annuli formed in the 2D section though the 3D
optical lattice at the center and (b) the corresponding radial
variation of density $\rho_i$ and superfluid density $\rho^s_i$. (c)
radial variation of $\rho_{i\sigma}$ and $\rho^s_{i,\sigma}$ and (d)
$N_\sigma(z)$ versus $z$ for $\sigma=\pm 1, 0$. Note that
$N_{-1}(z)=N_1(z)$. $N(z)=\sum_\sigma N_\sigma$. Here the optical
lattice parameters are taken to be $\mu/E_r=1$, $V_T/E_r=0.001$,
$V_0/E_r=14.5$ and $U_2/U_0=0.03$. }
 \label{fig:spin1}
\end{figure}

With the order parameters that we have defined in
Eq.~\ref{eq:opspin1} we can, first, obtain phase diagrams for the
spin-1 BH model for various values of $U_0$ and $U_2$; we refer the
reader to our earlier study~\cite{rvpspin1} for such phase diagrams
that include polar and ferromagnetic SF phases.  Here we use the
inhomogeneous MF theory, which we have developed above for the
spin-1 BH model, to obtain some illustrative results for
order-parameter profiles in a representative case that has a polar
superfluid.  In particular, we consider a simple-cubic lattice with
$70^3$ sites, $\mu/E_r=1$, $V_T/E_r=0.001$, $V_0/E_r=14.5$, and
$U_2/U_0=0.03$, the 2D planar section ${\cal P}_z$ for $z=0$ is
plotted in Fig.~(\ref{fig:spin1})(a). The radial variations of the
total on-site density of bosons $\rho_i=\sum_\sigma \rho_{i,\sigma}$
and total on-site superfluid density $\rho^s_i=\sum_\sigma
\rho^s_{i,\sigma}= \sum_\sigma \mid \psi_{i,\sigma}\mid^2$, as well
as the individual component of superfluid density $\mid
\psi_{i,\sigma}\mid^2$  are given in Figs.~(\ref{fig:spin1})(b) and
(c), respectively. From Fig.~\ref{fig:spin1}(b) it is evident that
this system has two well developed MI ($\rho=1$ and $2$) shells,
which are represented as regions with black and red respectively.
The most important result of
model~(\ref{eq:bhspin1})~\cite{rvpspin1} is that the superfluid
phase is polar for $U_2>0$ and, according to the symmetry
consideration, within our MF theory, the superfluid order parameters
take one of the two possible set of values; $\psi_{\pm 1}\ne 0$,
$\psi_0=0$ or $\psi_{\pm 1}= 0$, $\psi_0 \ne 0$.
Figure~(\ref{fig:spin1})(c) yields $\psi_\pm 1 >0$ and $\psi_0=0$ in
the superfluid phase confirming the polar nature of the phase.
Another important feature is that $\rho_{i,\pm 1} \ne \rho_{i,0}$ in
the polar superfluid phase. This leads to $N_{\pm 1}(z) \ne N_0(z)$
where $N_\sigma(z)$ is the total number of bosons with a spin
$\sigma$  in the 2D planar section ${\cal P}_z$ and is plotted in
Fig.~\ref{fig:spin1}(d) versus $z$.  Thus the determination of
$N_\sigma(z)$ experimentally can reveal these features and thus can
be used to confirm the  polar nature of the superfluid phase in
spin-1 bosons in optical lattice.

\begin{figure*}[htbp]
\centering \epsfig{file=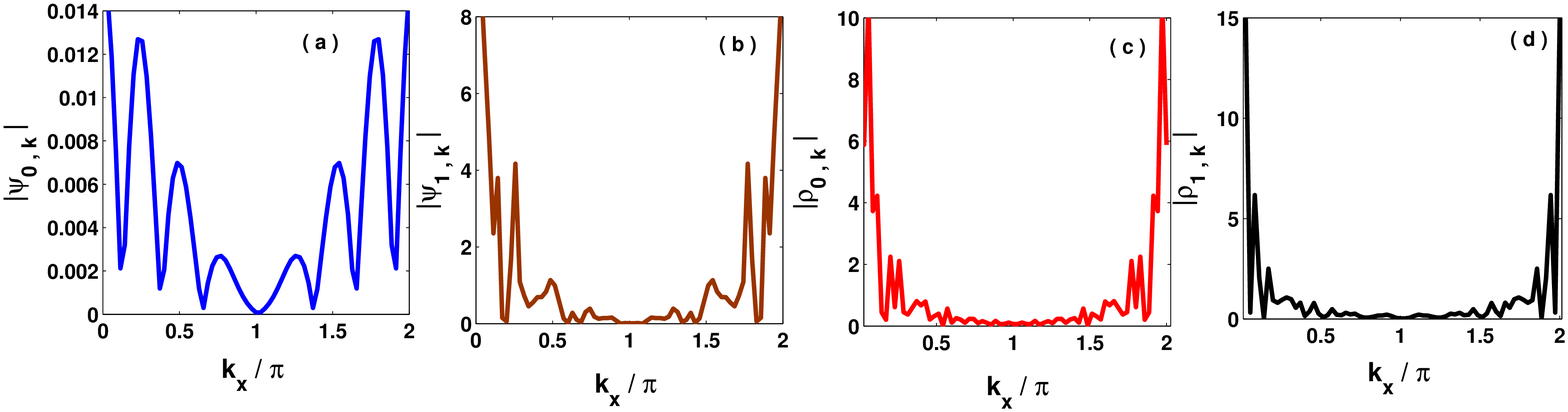,width=16cm,height=10cm}
\caption{(Color online) Illustrative plots of the moduli of the
one-dimensional Fourier transforms (a) $|\psi_{0,k}|$, (b)
$|\psi_{1,k}|$, (c) $|\rho_{0,k}|$, and (d) $|\rho_{1,k}|$ for the
profiles in Fig.~\ref{fig:spin1} (c).} \label{fig:spin1fourier}
\end{figure*}

In Fig.~\ref{fig:spin1fourier} we show moduli of the one-dimensional
Fourier transforms of the density and order-parameter profiles
in Fig.~\ref{fig:spin1} (c). It would be interesting to see if
such patterns can be obtained via time-of-flight measurements.

\begin{figure}[htbp]
\centering \epsfig{file=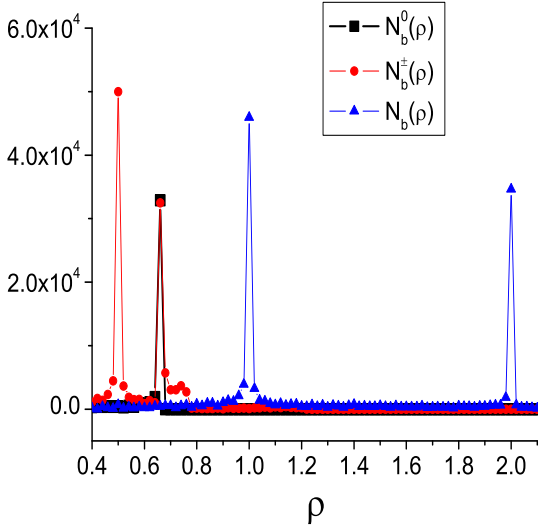,width=8cm,height=8cm}
\caption{(Color online)  Representative plots versus the density
$\rho$ of $N^\pm_b(\rho)$ (red circles), the total number of bosons,
at density $\rho$ and with $\sigma=\pm1$, $N^0_b(\rho)$ (black
squares), the total number of bosons, at density $\rho$ and with
$\sigma=0$, and their sum $N_b(\rho) =
N^+_b(\rho)+N^0_b(\rho)+N^-_b(\rho)$ (blue triangles); here all
parameters are as in Fig.~\ref{fig:spin1}.} \label{fig:spin1clock}
\end{figure}

In Fig.~\ref{fig:spin1clock} we show a representative plot of the analog of
Fig.~\ref{fig:freq} for the spin-1 BH model with parameter values as in
Fig.~\ref{fig:spin1}; thus, there are two well-developed MI shells. Here
$N^\pm_b(\rho)$ denotes the total number of bosons, at density $\rho$, and
with $\sigma=\pm1$; similarly, $N^0_b(\rho)$ is the total number of bosons,
at density $\rho$, and with $\sigma=0$; and is $N_b(\rho)$ is the total
number of bosons at density $\rho$. For the peak in $N_b(\rho)$,
in the vicinity of $\rho=1$, only bosons with $\sigma=\pm 1$ contribute; but,
for the one near $\rho=2$, all three components contribute equally.  This
result, which is also implicit Fig.~\ref{fig:spin1} (c), should be verifiable
in atomic-clock-shift experiments of the type that have been carried out for
spinless bosons~\cite{campbell}.

We have noted in earlier work~\cite{rvpspin1} that our mean-field theory does
not account for order parameters that distinguish between different spin
orderings, which have been studied~\cite{mimag} in the limit $U_0\to\infty$,
in the MI phases in spin-1 BH models. The exploration of such orderings lies
beyond the scope of the present study.

\section{Conclusions}

We have carried out a comprehensive study of Mott insulator and
superfluid shells in Bose-Hubbard models for bosons in optical
lattices with harmonic traps by using an intuitively appealing
inhomogeneous mean field theory that has been used earlier to
understand the Bose-glass phase~\cite{sheshprl}. Our inhomogeneous
mean-field theory quantitatively agrees with QMC simulations.
Furthermore, it is numerically less intensive than QMC simulations;
thus, we are able to perform calculation on experimentally
realistic, large 3D systems and explore a wide range of parameter
values. We can calculate in-trap density profiles that agree
qualitatively with experiments~\cite{bloch}; and we show how to
obtain the phase diagram of the homogeneous Bose Hubbard model from
such in-trap density profiles. Our results are also of direct
relevance to recent atomic-clock-shift experiments~\cite{campbell}
as we have described above.  Finally we have generalized  our
inhomogeneous mean-field theory to BH models with two species of
bosons or a spin-1 BH model with harmonic traps.  With two species
of bosons we obtain rich phase diagrams with a variety of SF and MI
phases and associated shells, when we include a quadratic confining
potential; we also obtain in-trap density distributions that show
plateaux as in the single-species case. For the spin-1 BH model we
show, in a representative case, that the system can display
alternating shells of polar SF~\cite{rvpspin1} and MI phases; and we
make  interesting predictions for atomic-clock-shift experiments. We
hope our results will stimulate more experiments on such systems of
bosons in optical lattices.  Our inhomogeneous mean-field theory can
also be generalized to study the extended Bose-Hubbard model as we
report elsewhere~\cite{kurdestany}.

Though other groups~\cite{bergkvist,pollet,demarco05,mitra08,spiel,kato08}
have studied such shell structure theoretically, they have not obtained the
quantitative agreement with quantum Monte Carlo (QMC)
simulations~\cite{wessel} that we obtain, except in one
dimension~\cite{batrouni08}. Furthermore, there have been some investigations
of the BH model with a harmonic trap potential; these use mean-field
theory~\cite{spiel,ozaki09} and, in addition, a local-density approximation
(LDA), which assumes that the properties of a system with finite confining
potential at a particular location are identical to those of a uniform system
with the value of the local chemical potential at that location.
%example \begin{eqnarray}\label{lda} \rho_i^s(U,\mu,V_T)&=&
%\rho_i^s(U,\mu_i,0) \nonumber \\ \rho_i(U,\mu,V_T)&=&\rho_i(U,\mu_i,0)
%\end{eqnarray} where $\mu_i=\mu-V_TR^2_i$.
This approximation leads to a decoupling of each site from its neighbor; it
is equivalent to assuming $\phi_i=\psi_i$ in Eq.~\ref{eq:mfh0} and a
minimization of the ground-state energy for each site separately. In our
inhomogeneous mean-field theory, we do not make this additional LDA
assumption; and the minimization of the ground state energy is done over the
entire set of ${\psi_i}$.  If we compare these two approaches for the
single-species BH model, we find that the difference is negligible in SF
regions, but discrepancies exist at SF-MI interfaces; this has been reported
in other models~\cite{bou_lda} also.

\section{ACKNOWLEDGMENT}

We thank H.R. Krishnamurthy for discussions and DST,
UGC, and CSIR (India) for support. One of us (RVP) thanks the
Jawaharlal Nehru Centre for Advanced Scientific Research and the
Indian Institute of Science, Bangalore for hospitality.

\end{document}